\documentclass[twocolumn,english]{IEEEtran}
\usepackage[T1]{fontenc}
\usepackage{babel}
\usepackage{array}
\usepackage{multirow}
\usepackage{amsthm}
\usepackage{amsmath}
\usepackage{amssymb}
\usepackage{graphicx}
\usepackage{epstopdf}
\usepackage[unicode=true,
 bookmarks=true,bookmarksnumbered=true,bookmarksopen=true,bookmarksopenlevel=1,
 breaklinks=false,pdfborder={0 0 0},backref=false,colorlinks=false]
 {hyperref}
\hypersetup{pdftitle={Your Title},
 pdfauthor={Your Name},
 pdfpagelayout=OneColumn, pdfnewwindow=true, pdfstartview=XYZ, plainpages=false}

\makeatletter

\providecommand{\tabularnewline}{\\}

 \let\oldforeign@language\foreign@language
 \DeclareRobustCommand{\foreign@language}[1]{%
   \lowercase{\oldforeign@language{#1}}}
\theoremstyle{plain}
\newtheorem{thm}{\protect\theoremname}

\ifCLASSOPTIONcompsoc
\usepackage[caption=false,font=normalsize,labelfont=sf,textfont=sf]{subfig}
\else
\usepackage[caption=false,font=footnotesize]{subfig}
\fi

\@ifundefined{showcaptionsetup}{}{%
 \PassOptionsToPackage{caption=false}{subfig}}
\usepackage{subfig}
\makeatother

\providecommand{\theoremname}{Theorem}

\begin{document}

\title{Generic Evolving Self-Organizing Neuro-Fuzzy Control of Bio-inspired
Unmanned Aerial Vehicles}

\author{Md Meftahul~Ferdaus,~\IEEEmembership{Student Member,~IEEE,} Mahardhika
~Pratama,~\IEEEmembership{Member,$\:$IEEE,} Sreenatha G. ~Anavatti,~Matthew
A. Garratt,~and~Yongping Pan%
\thanks{Md Meftahul~Ferdaus is with~the School of Engineering and Information
Technology, University of New South Wales at the Australian Defence
Force Academy, Canberra, ACT 2612, Australia, e-mail: \protect\href{mailto:m.ferdaus@unsw.edu.au}{m.ferdaus@unsw.edu.au}.%
}%
\thanks{Mahardhika Pratama~is with the School of Computer Science and Engineering,
Nanyang Technological University, Singapore, 639798, Singapore, e-mail:
\protect\href{mailto:mpratama@ntu.edu.sg}{mpratama@ntu.edu.sg}.%
}%
\thanks{Sreenatha G. ~Anavatti~ is with the School of Engineering and Information
Technology, University of New South Wales at the Australian Defence
Force Academy, Canberra, ACT 2612, Australia, e-mail: \protect\href{mailto:s.anavatti@adfa.edu.au}{s.anavatti@adfa.edu.au}.%
}%
\thanks{Matthew A. Garratt~ is with the School of Engineering and Information
Technology, University of New South Wales at the Australian Defence
Force Academy, Canberra, ACT 2612, Australia, e-mail: \protect\href{mailto:M.Garratt@adfa.edu.au}{M.Garratt@adfa.edu.au}.%
}%
\thanks{Yongping Pan~ is with the Department of Biomedical Engineering, National
University of Singapore, Singapore, e-mail: \protect\href{mailto:biepany@nus.edu.sg}{biepany@nus.edu.sg}.%
}}

\IEEEspecialpapernotice{}

\IEEEaftertitletext{}

\markboth{IEEE Transactions on Fuzzy Systems}{M M Ferdaus \MakeLowercase{\emph{et al.}}:
Online Control of Autonomous Vehicles based on Generic Controller}
\maketitle
\begin{abstract}
At recent times, with the incremental demand of the fully autonomous
system, a huge research interest is observed in learning machine based
intelligent, self-organizing, and evolving controller. In this work,
a new evolving and self-organizing controller namely Generic-controller
(G-controller) is proposed. The G-controller that works in the fully
online mode with very minor expert domain knowledge is developed by
incorporating the sliding model control (SMC) theory based learning
algorithm with an advanced incremental learning machine namely Generic
Evolving Neuro-Fuzzy Inference System (GENEFIS). The controller starts
operating from scratch with an empty set of fuzzy rules, and therefore,
no offline training is required. To cope with the plant's vulnerable
behavior, the controller can add, or prune the rules on demand. Control
law and adaptation laws for the consequents are derived from the SMC
algorithm to establish a stable closed-loop system, where the stability
of the G-controller is guaranteed using the Lyapunov function. The
uniform asymptotic convergence of tracking error to zero is witnessed
through the implication of an auxiliary robustifying control term.
In addition, the implementation of the multivariate Gaussian function
helps the controller to handle the non-axis parallel data from the
plant and consequently enhances the robustness against the uncertainties
and environmental perturbations. Finally, the controller's performance
has been evaluated by observing the tracking performance in controlling
simulated plants of unmanned aerial vehicle namely bio-inspired flapping
wing micro air vehicle (BIFW MAV) and hexacopter for a variety of
trajectories.\end{abstract}

\section{Introduction\label{sec:Introduction}}

\IEEEPARstart{ I}{n autonomous} unmanned vehicle systems, to obtain
an accurate first principle based model is considerably arduous due
to their highly non-linear and over-actuated or under-actuated behavior.
Besides, various uncertainty factor like impreciseness in the obtained
data from sensors, induced noise by the sensors, outdoor environmental
uncertainties like wind gust, motor degradation, etc. are strenuous
or impossible to integrate into the first principle models. In such
circumstances, approaches without the necessity of accurate mathematical
models of the system under control, are much appreciated. Being a
model-free approach, the Neural Network (NN) and Fuzzy Logic system
(FLS) based controllers have been successfully implemented in many
control applications \cite{gomi1993neural,liu2015adaptive,noriega1998direct,takagi1993fuzzy}
over the past few years. Recently, systems with an amalgamation of
FLS and NN namely Fuzzy Neural Network (FNN) based controllers are
becoming popular in a variety of engineering application, such as
in controlling robot manipulator \cite{wai2004intelligent}, anti-lock
braking systems \cite{peric2016quasi}, temperature control \cite{lin2004ga},
controlling motor drive \cite{lin2001self}, direct current converter
\cite{wai2012adaptive} etc.

To develop an FLS, NN, or an FNN controller with a better accuracy,
is challenging. There exist numerous methods to develop FLS, NN, or
FNN controllers \cite{pan2015peaking,pan2017biomimetic,pan2017composite}.
When the system dynamics and various characteristics of a plant to
be controlled are known, then the information is utilized to train
the controller. It constructs a fixed-structured controller with a
certain number of rules, membership functions, neurons, and layers.
Due to the fixed structure, these controllers cannot cope with changing
plant dynamics. Similarly, during online control application, the
plant dynamics and other system information are not readily available.
Therefore, the fixed-structured FLS, NN, or FNN controller become
unreliable. Furthermore, the characteristic of real-world plants is
non-stationary. This should not be handled by a fixed-size controller.
To handle uncertainties in control system, researchers have tried
to combine the FLS, NN, FNN system with sliding mode control (SMC)
\cite{sun2011neural}, $H_{\infty}$ control, back-stepping, etc.
Such amalgamation empowers the FLS, NN, FNN controller with the feature
of tuning parameters, which provides a more robust and adaptive control
structure. It assists them to mitigate the adverse effects of various
uncertainties and perturbations. However, such adaptive FNN control
structures are not able to evolve their structures by adding or pruning
rules. It forces the controller to determine the number of rules a
priori, where a selection of few fuzzy rules may hinder to achieve
adequate and desired control performance. On the other hand, consideration
of too many rules may create complexity, or make it impossible to
implement in real time.

The problems mentioned above can be circumvented by implementing evolving
FLS, NN, FNN controllers that can evolve their structure by adding,
or deleting rules through self-organizing techniques. In recent time,
researchers are trying to develop evolving FLS, NN, FNN controllers
by following various approaches to add or delete the rules \cite{pratama2014panfis}.
In the rule growing mechanism in \cite{gao2003online}, $\epsilon-$completeness
and system error was measured to add a rule, and the concept of error
reduction ratio (ERR) was utilized to prune a rule. Therefore, their
developed self-organizing FNN controller can adapt both the parameters
and the structure. Nevertheless, the computation of larger matrix
in each step, and storing of all previous input-output data makes
their proposed technique computationally costly. Sometimes, the spatial
proximity between a particular data-point and all other points, namely
potential is used to add or replace a rule in an evolving fuzzy controller
as described in \cite{angelov2004fuzzy}. However, the controller
proposed in \cite{angelov2004fuzzy} cannot prune any rules, which
makes them computationally costly in real-time control, especially
in case of controlling highly nonlinear over-actuated complex systems.
Researchers have also tried in \cite{lin2009self} to employ a max-min
method to add and prune layers, and error backpropagation to tune
evolving neuro-fuzzy controller's parameters like weight, center,
and width of Gaussian membership function. Nonetheless, the controller
needed some knowledge about the plant like the bounded nominal matrix,
which may not be known during control operation. 

Besides, an evolving TS fuzzy controller can be developed by combining
it with a fixed TS fuzzy FLS as exposed in \cite{de2007evolving},
where the evolving TS FLS worked online, and no previous information
was needed. Though such design of an evolving controller is simple,
it requires some information about the plant due to the utilization
of static TS fuzzy system. Researchers have tried to develop a cloud-based
evolving fuzzy controller in \cite{vskrjanc2014robust}. The advantage
of their cloud-based evolving fuzzy controller was the parameter-free
antecedent part of a rule, whereas the consequent part was expressed
in the form of a PID controller. The three parameters of their PID
like evolving controller were tuned online through Lyapunov theory
based stable adaptation law. The controller can add rules or clouds
by measuring local data density. Their evolving structure helped them
to track the desired trajectory properly. However, they were not able
to delete rules, or clouds. An evolving fuzzy controller is possible
to develop by using Fuzzy granular computation in explained \cite{leite2015evolving},
which can model and control an unknown non-stationary system in online
without any experts knowledge. A fuzzy Lyapunov function was utilized
to evaluate the closed-loop stability of their controller. Data uncertainty
was handled and incorporated into knowledge domain by Fuzzy granular
computation. However, diversity in data among different granules,
and unmeasurable state variables was not considered.

Furthermore, an evolving neuro-fuzzy controller can also be developed
by utilizing model predictive control methods as described in \cite{han2013real,han2014nonlinear,han2016nonlinear}.
A radial basis function neural network based self-organizing controller
was developed in \cite{han2013real} where the evolving neuro-fuzzy
system was utilized to bound the modeling error uniformly for nonlinear
systems. Besides, a fast gradient method was employed to minimize
the computational cost of their online evolving model predictive control
method. The Lyapunov theory demonstrated the closed-loop stability
of their control method. However, their controller was not implemented
in real-time application. 

Until now, all the discussed evolving controllers have utilized univariate
Gaussian function, which does not expose the scale-invariant property.
Besides, they are not effective in dealing with non-axis parallel
data distribution. To mitigate these shortcomings, we utilized a multivariate
Gaussian functions based incremental learning machine algorithm called
generic evolving neuro-fuzzy inference system (GENEFIS) \cite{pratama2014genefis}.
In this work, GENEFIS is amalgamated with SMC technique to develop
a self-evolving controller namely G-controller. In this work, main
features of the proposed G-controller are as follows:
\begin{enumerate}
\item The controller's performance does not depends upon the plant dynamics
or any other features of the plant.
\item No previous information or off-line training is required. Thus, the
controller starts self-construction from scratch with only one rule
at the beginning, and then it adds, deletes, or merge rules to follow
the desired trajectory. Besides, the application of a fast kernel-based
metric approach helps to capture fuzzy set and rule level redundancy.
\item Integration of the Generalized Adaptive Resonance Theory+ (GART+)
helps to upgrade the premise parameters with respect to input data
distributions, and utilization of multivariate Gaussian function aids
the controller to handle a variety of data generated from the sudden
change in plants, or from uncertainties, environmental perturbations.
\item Adaptation law for the GENEFIS based G-controller's consequent parameters
are derived from the SMC learning theory, which confirms a stable
closed-loop control system. A robustifying auxiliary control term
ensures uniform asymptotic convergence of tracking error to zero.
Finally, the stability of the G-controller is proved using the Lyapunov
function.
\item Successful evaluation of the proposed G-controller through implementing
it into the simulated BIFW MAV, and hexacopter plant.
\end{enumerate}
The above-mentioned characteristics of the proposed self-evolving
G-controller make them an appropriate candidate to control autonomous
vehicles like BIFW MAV, quadcopter, hexacopter, octocopter with better
accuracy than a stand-alone first principle based controller. Furthermore,
the whole controller is developed using C programming language to
make it compatible with all types of hardware, where their implementation
is made publicly available in \cite{mpratamaweb}. In the next subsection
\ref{sub:Related-Work}, the recent implementation of FLS, NN, FNN
based controller in autonomous vehicles are summarized.

\subsection{Related Work\label{sub:Related-Work}}

Autonomous vehicles express a complex mathematical model, where the
incorporation of various uncertainties is very difficult, or not possible
in some cases. The situation becomes worse, when those vehicles are
highly nonlinear, under-actuated, or over-actuated like bio-inspired
flapping wing micro air vehicle (BIFW MAV), quadcopter, hexacopter,
octocopter unmanned aerial vehicles \cite{pratama2013online}. It
is due to the associated uncertainties, and rarely available information
or no information at all about the autonomous vehicles. As a consequence,
in recent time research interest is increasing in employing model-free
adaptive and self-evolving fuzzy and neuro-fuzzy controller to autonomous
vehicles. A self-organizing adaptive fuzzy controller was developed
in \cite{wang2016adaptive} to control a complex surface vehicle with
uncertainties, and external perturbation. A robustifying auxiliary
control term contributed to ensure closed-loop system stability of
their controller. From simulation studies, better performance was
observed from their fixed adaptive control techniques. However, a
higher number of rules were generated to achieve desired tracking
accuracy. In \cite{ferdaus2017fuzzy}, an adaptive fuzzy-PID controller
was developed to control the attitude and altitude of a quadrotor.
Though the TS-fuzzy system can approximate the plant's dynamics with
uncertainties and unknown disturbances, their dependency of a static
PID controller affected their performance to cope with sudden changes
in plant dynamics.

An evolving fuzzy reference controller was proposed in \cite{dovvzan2014towards},
which can partition the input-output space. It was able to identify
the local controllers gain. The stability of their controller is yet
to prove. In \cite{zdevsar2014self}, self-tuning of a two-degrees-of-freedom
control algorithm was proposed based on an evolving fuzzy model. Their
control algorithm consisted of a feedforward, and feedback loop. The
feedforward part was derived from inversion of TS fuzzy model, which
helped the system to follow the trajectory closely. An evolving fuzzy
system with the ability to add, prune, merge, or split the clusters
was utilized in the feedback loop of their controller. Nonetheless,
their controller was applicable for single-input-single-output (SISO)
systems only. In \cite{cara2013new}, a Taylor Series Neuro Fuzzy
(TaSe-NF) model based online self-evolving neuro fuzzy controller
is developed. The local online learning of consequent parameters provided
them a proper control mechanism, and the self-evolving structure provided
satisfactory performance with minimum rules and no prior information
about the plant. However, their controller's stability yet to be proved.

A direct evolving neural controller was developed in \cite{han2015direct}
by implementing a self-constructing radial basis function neural network
(RBFNN), which can add or prune neurons online to adapt to plant condition.
The closed-loop stability of their controller was guaranteed by Lyapunov
stability theory. A self-organizing fuzzy controller is proposed in
\cite{cara2011new}, where the consequents are adapted at each selected
topology, and membership functions were added online. After analysing
the complete operating region of the plant the topology was modified,
which raised their controllers robustness. However, their controller
had a dependency of plants sampling period, which made them impractical
in implementing real embedded UAV plants.

An interval type-2 fuzzy neural network based gain adaptive sliding
mode controller was developed in \cite{li2013gain}to control the
attitude micro aerial vehicle. The sliding mode gain adaptive law
was utilized to tune the parameters of the interval type-2 fuzzy neural
network online. Better tracking accuracy was observed than the conventional
adaptive sliding mode controller. However, the fuzzy controller can
only tune its parameters, and unable to evolve their structure by
adding or deleting rules. A self-organizing interval type-2 fuzzy
neural network (SOIT2FNN) based controller was developed in \cite{chen2011robust},
and combined with a PD controller to control the attitude of a quadcopter
Micro Aerial Vehicle (MAV), where the SOIT2FNN was able to learn the
inverse model of quadcopter online. Besides, it could minimize the
model errors and external disturbances. Nonetheless, due to their
dependency on static PD controller, they were affected by sudden changes
in plant dynamics. 

To deal with the uncertainties exists in the nonlinear complex FW
MAV model, a radial basis function neural network (RBFNN) was utilized
in \cite{he2016trajectory}. The stability of their closed loop control
system was verified through Lyapunov stability theorem. By selecting
appropriate control variables, a good trajectory tracking performance
was observed from their simulation results. However, their control
technique was not self-evolving in nature. A fuzzy controller namely
hybrid adaptive fuzzy controller (HAFC) was developed in {[}14{]}
to control a dragonfly-like FW MAV. Their HAFC was able to tune its
parameters by using the hybrid adaptive law to minimize the tracking
error. The simulation results of the HAFC were compared with a direct
adaptive (DA) method based fuzzy controller (DAFC). Better tracking
performance was observed from the HAFC than the DAFC. However, their
controller was constructed by the batch learning process, therefore
static in nature. A Spiking Neural Network (SNN) based controller
was developed for an FW MAV called RoboBee in \cite{clawson2016spiking},
where they had utilized a reward-modulated Hebbian plasticity mechanism
to adopt a leaky integrate-and-fire spiking neural network in flight.
In \cite{suresh2012line}, an online self-organizing neural controller
was combined with a conventional controller to track the desired trajectory
of a simulated helicopter plant, while performing highly nonlinear
maneuvers. In their dynamic radial basis function network, a Lyapunov
based adaptation law was integrated with the neuron growing and pruning
mechanism, which ensured the closed-loop stability and desired tracking
accuracy. A self-constructing neuro-fuzzy controller was proposed
in \cite{dong2016self} to track the desired attitude of a quadcopter,
which can add, and delete rules online. Integration of a projection
based adaptation law handled the drift, and singularity problem in
antecedent parameters. However, due to their rule growing mechanism
comparatively higher number of rules were generated online, which
made them computationally costly. 

To adapt the consequent parameters of various self-evolving neuro-fuzzy
controllers, sometimes gradient-based algorithms are utilized. These
gradient-based algorithms perform better only with those plants, where
a slow variation in their dynamics is observed. Besides, some gradient-based
algorithms namely dynamic backpropagation includes partial derivatives,
which may slow down the convergence speed, especially in case of complex
search space. In addition, the tuning procedure may be trapped into
a local minimum in backpropagation methods as explained in \cite{topalov2001online}.
As a solution to these problems, evolutionary approaches are proposed
in \cite{topalov1996fast}. However, the stability of such controller
is questionable. To overcome these issues, sliding mode control (SMC)
theory based algorithms were proposed to tune the consequent parameter
of various FLS, NN, FNNs in \cite{parma1998sliding,yu2004fuzzy,cascella2005adaptive,efe2000sliding}.
An SMC theory based adaptive and online neuro-fuzzy-PID controller
was developed in \cite{kayacan2013adaptive}, which exhibited an improved
performance than standalone PID controller. Since the adaptation laws
are derived from the PID controller part, without the PID controller,
the neuro-fuzzy controller was unable to tune its parameters.

To mitigate the shortcomings of the existing self-evolving neuro-fuzzy
controller, in this work a novel self-evolving controller namely G-controller
is developed in guiding various autonomous vehicles. The evolving
architecture of the controller is inspired by the incremental learning
algorithm called GENEFIS \cite{pratama2014genefis}, and the consequent
parameters are adapted utilizing SMC theory based learning algorithm.
Unlike the evolving neuro-fuzzy controllers discussed in the literature,
the integration of GART+ with a four-stage checking to evolve rules
provide a very quick response, and reduce the computational complexity
by pruning unnecessary rules. Instead of predefined values, the sliding
parameters in the SMC theory is also self-organizing in this work.
To the best of our knowledge, such evolving sliding parameters are
never utilized before in any of the existing evolving neuro-fuzzy
controllers.

The organization of rest of this paper is as follows. Section \ref{sec:Problem-Statement}
describes the problem associated with two autonomous vehicles namely
BIFW MAV, and hexacopter in tracking the trajectory accurately. The
self-evolving architecture of the GENEFIS based G-controller is explained
in Section \ref{sec:Architecture-of-Self-Evolving}. Section \ref{sec:SMC-Theory-based-Adaptation}
represents SMC learning algorithm based adaptation of the proposed
G-controller. The results are summarized, and analysed in Section
\ref{sec:Results-and-Discussion}. Finally, the paper ends with the
concluding remarks encompassed in Section \ref{sec:Conclusions}.

\section{Problem Statement\label{sec:Problem-Statement}}

Modelling and control of BIFW MAV is one of the latest research topics
related to the field of autonomous Unmanned Aerial Vehicles (UAVs).
BIFW MAV exhibits some advanced characteristics like fast flight,
vertical take-off and landing, hovering, and quick turn, and enhanced
manoeuvrability when compared to similar sized fixed and rotary wing
UAVs. To observe these features from a BIFW MAV, an advanced control
mechanism is a must need. To take the challenge, our proposed G-controller
is implemented in a bio-inspired (BI) dragonfly-like four wings simulated
MAV plant developed by the Defence Science and Technology Group (DSTG)
as described in \cite{kok2015low}. The simulated BIFW MAV plant saves
the time and expenses to set-up for experimental flight test. The
high level architecture of the BIFW MAV flight simulator is exposed
in Fig. \ref{fig:High-level-architecture_flap}. 

\begin{center}
\begin{figure*}[t]
\begin{centering}
\includegraphics[scale=0.28]{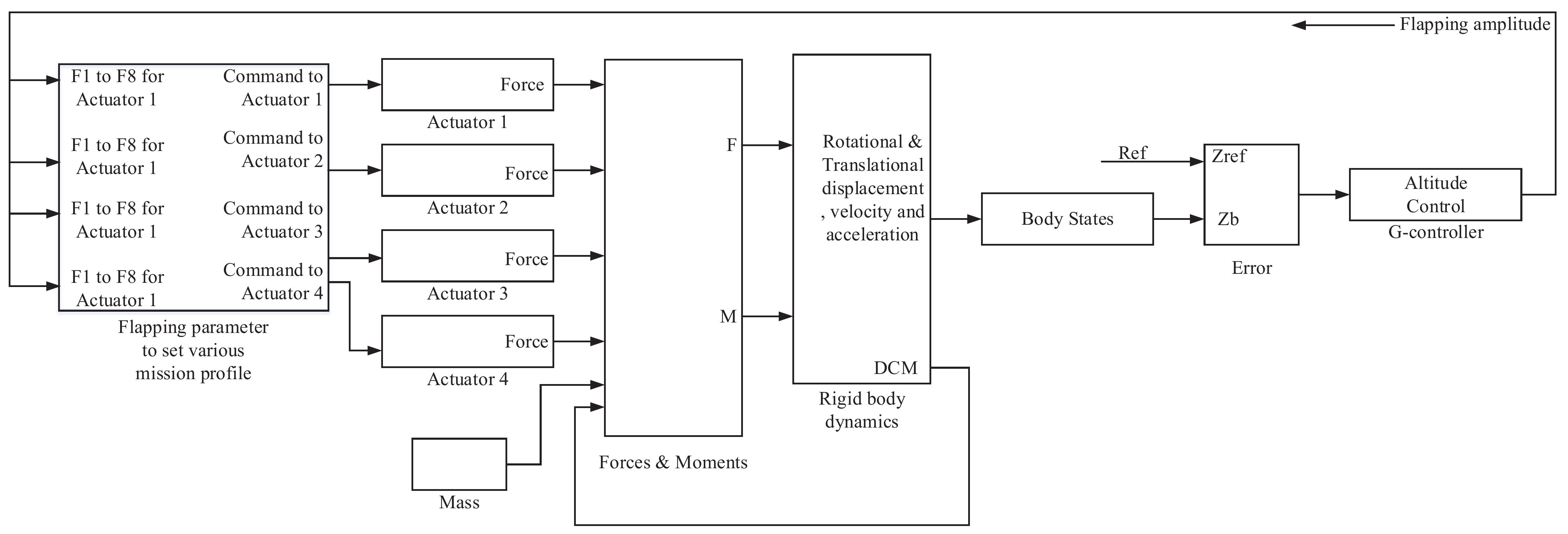}
\par\end{centering}

\caption{High level architecture of the over-actuated FWMAV plant\label{fig:High-level-architecture_flap}}

\end{figure*}

\par\end{center}

In the flight simulator, four actuators are denoted by 'Actuator 1',
'Actuator 2', 'Actuator 3', and 'Actuator 4' as shown in Fig. \ref{fig:High-level-architecture_flap}.
These actuators are utilized to control four wings of the BIFW MAV.
Each actuator is controlled by eight (8) flapping parameters indicated
by F1 to F8 in 'flapping parameters' block, where F1 indicates the
stroke plane angle (in rad), F2 is the flapping frequency (in Hz),
F3 represents the flapping amplitude (in rad), F4 presents the mean
angle of attack (in rad), F5 is the amplitude of pitching oscillation
(in rad), F6 is the phase difference between the pitching and plunging
motion, F7 is the time step (in sec), F8 is the kappa set as zero
in the plant. By altering these flapping parameters, from this flight
simulator various mission profile or manoeuvrability such as take-off,
rolling, pitching, and yawing of an BIFW MAV can be analysed. To find
the dominant flapping parameter, a parametric analysis is accomplished
in this work. It is found that the flapping amplitude is the most
dominant among the eight parameters to control the BIFW MAV. By tuning
the flapping amplitude the take-off, rolling, and pitching motion
is observed. Only for the yawing motion flapping phase needs to be
tuned. Tuning of these parameters and their effect on various manoeuvring
is summarized in\textbf{ }TABLE \ref{tab:Effects-of-flapping}. 

\begin{center}
\begin{table}[t]
\caption{Effects of flapping parameters in different manoeuvring of FW MAV($\phi_{0}$:
Flapping amplitude(degree); $\Psi$: Phase(degree))\label{tab:Effects-of-flapping}}

\centering{}%
\begin{tabular}{|c|l|}
\hline 
\textbf{Actuators with corresponding flapping parameter} & \textbf{Action}\tabularnewline
\hline 
Actuator: 1, 2, 3, 4; $\phi_{0}$: 90 & Take-off\tabularnewline
\hline 
Actuator: 1, 2; $\phi_{0}$: 90 and Actuator: 3, 4; $\phi_{0}$: 60 & Roll-right\tabularnewline
\hline 
Actuator: 1, 2; $\phi_{0}$: 60 and Actuator: 3, 4; $\phi_{0}$: 90 & Roll-left\tabularnewline
\hline 
Actuator: 1, 3; $\phi_{0}$: 90 and Actuator: 2, 4; $\phi_{0}$: 60 & Pitch-up\tabularnewline
\hline 
Actuator: 1, 3; $\phi_{0}$: 60 and Actuator: 2, 4; $\phi_{0}$: 90 & Pitch-down\tabularnewline
\hline 
Actuator: 1, 4; $\Psi$: 90 and Actuator: 2, 3; $\Psi$: 60 & Yaw-right\tabularnewline
\hline 
Actuator: 1, 4; $\Psi$: 60 and Actuator: 2, 3; $\Psi$: 90 & Yaw-left\tabularnewline
\hline 
\end{tabular}
\end{table}

\par\end{center}

The force generated by each actuator $(F_{r_{i}})$ is supplied to
the 'Forces \& Moments' block of the Fig. \ref{fig:High-level-architecture_flap}.
Besides the four forces from four actuators ($F_{r_{1}}$, $F_{r_{2}}$,
$F_{r_{3}}$, $F_{r_{4}}$), the mass and Direction Cosine Matrix
(DCM) is also fed to the 'Forces \& Moments' block. This block supplies
the required force (F) and moment (M) to the rigid body dynamics of
the simulator based on the relative airflow acting on each wing and
the commanded actuator speed. The wing kinematics were modelled utilizing
the derivation explained in \cite{wang2004role,wang2005dissecting,wang2007effect},
where it was considered that the wing was flapping in an inclined
stroke plane with a certain angle. In the simulator, to express the
MAV's flapping profile the flapping angle ($\phi$) is presented in
a sinusoidal form as follows:

\begin{equation}
\phi(t)=\phi_{a}\text{cos}(\pi ft)\label{eq:1}
\end{equation}
 where $\phi_{a}$ is the flapping amplitude in radian, $f$ is the
flapping frequency in Hz, $t$ is the time is second. Besides, the
angle of attack ($A_{aoa}$) can be expressed as: 
\begin{equation}
A_{aoa}=A_{mn}-A_{p}\text{sin}(\omega dt+\Psi)\label{eq:2}
\end{equation}
 where $A_{mn}$ is mean angle of attack in radian, $A_{p}$ is amplitude
of pitching oscillation in radian, $dt$ is time step in seconds,
and $\Psi$ is the phase difference between the pitching and plunging
motion. All the four wings of the BIFW MAV follows the same flapping
profile.

Now in the 'Forces \& Moments' block $M$ is actually a summation
of four momentums of four different actuators, which can be expressed
as: 
\begin{equation}
M=M_{1}+M_{2}+M_{3}+M_{4}
\end{equation}

Each of these moments $(M_{1},\; M_{2},\; M_{3},\; M_{4})$ is actually
the torque induced in each wing based on the relative airflow acting
on each of them. The individual momentum of each wing can be expressed
as: 
\begin{equation}
M_{1}=F_{r_{1}}\times(CG-CP_{1})
\end{equation}
\begin{equation}
M_{2}=F_{r_{2}}\times(CG-CP_{2})
\end{equation}
\begin{equation}
M_{3}=F_{r_{3}}\times(CG-CP_{3})
\end{equation}
\begin{equation}
M_{4}=F_{r_{4}}\times(CG-CP_{4})
\end{equation}
where, $CG=[0\;0\;0];$ and $CP_{1}=[0.08\;0.05\;0]$; $CP_{2}=[0.08\;0.05\;0]$;
$CP_{3}=[0.08\;-0.05\;0];$ $CP_{4}=[-0.08\;-0.05\;0]$; and $\times$
is presenting $(3\times3$) cross product. Each of the forces $(F_{r_{i}})$
are three dimensional where the elements are $[F_{xr}\; F_{yr}\; F_{zr}]$.
On the other hand, $F$ is actually a summation of four forces of
four different actuators along with the gravitational force, which
can be expressed as: 
\begin{equation}
F=F_{r_{1}}+F_{r_{2}}+F_{r_{3}}+F_{r_{4}}+(mg\times DCM)
\end{equation}
 where, $m$ is the mass, $g$ is the acceleration due to gravity.
Finally, in the 'Rigid Body Dynamics' block the forces and moments
are converted into the body coordinate system, and all the required
body states like three dimentional rotational displacements ($\mbox{roll}(\phi),$$\:$$\mbox{pitch}(\theta),$$\:$$\mbox{yaw}(\psi$)),
velocities$\,$($\omega_{bx},\:\omega_{by},\:\omega_{bz}$), and accelerations$\,$($\alpha_{bx}=\frac{d\omega_{bx}}{dt},\:\alpha_{by}=\frac{d\omega_{by}}{dt},\:\alpha_{bz}=\frac{d\omega_{bz}}{dt}$)
and translational displacements ($X_{b},$ $Y_{b}$, $Z_{b}$), velocities
($v_{bx},\: v_{by},\: v_{bz}$), and accelerations ($a_{bx}=\frac{dv_{bx}}{dt},\: a_{by}=\frac{dv_{by}}{dt},\: a_{bz}=\frac{dv_{bz}}{dt}$)
are obtained, and the MAV states are updated. The complete mathematical
model of the BIFW MAV dynamics is written in C code and MATLAB m file,
where the C code is called using Simulink S-function. The complete
BIFW MAV simulator with the necessary files describing the dynamics
is made publicly available in \cite{mpratamaweb}. In addition, the
development of hardware construction of the above described BIFW MAV
plant is going on at this moment by the DSTG group. Their developed
flapping wing-actuation system is shown in the supplementary document.
After successful completion of the construction, our proposed G-controller
will be implemented in hardware, since the G-controller is compatible
with their hardware. 

Besides the BIFW MAV, another autonomous vehicle namely hexacopter,
a rotary wing UAV is considered in this work. The simulated hexacopter
plant is developed by UAV laboratory of the UNSW at the Australian
Defence Force Academy. The model is of medium fidelity and contains
both full 6 degrees of freedom (DOF) rigid body dynamics and non-linear
aerodynamics. The hexacopter simulated plant introduces two extra
degrees of freedom which are obtained by shifting two masses using
two aircraft servos with each mass sliding along its own rail aligned
in longitudinal and lateral directions respectively, which makes the
plant an over-actuated system. The top-level diagram of this over-actuated
simulated plant is exhibited in Fig. \ref{fig:Top-level-diagram-hexacopter}.

\begin{center}
\begin{figure*}[tbh]
\begin{centering}
\includegraphics[scale=0.23]{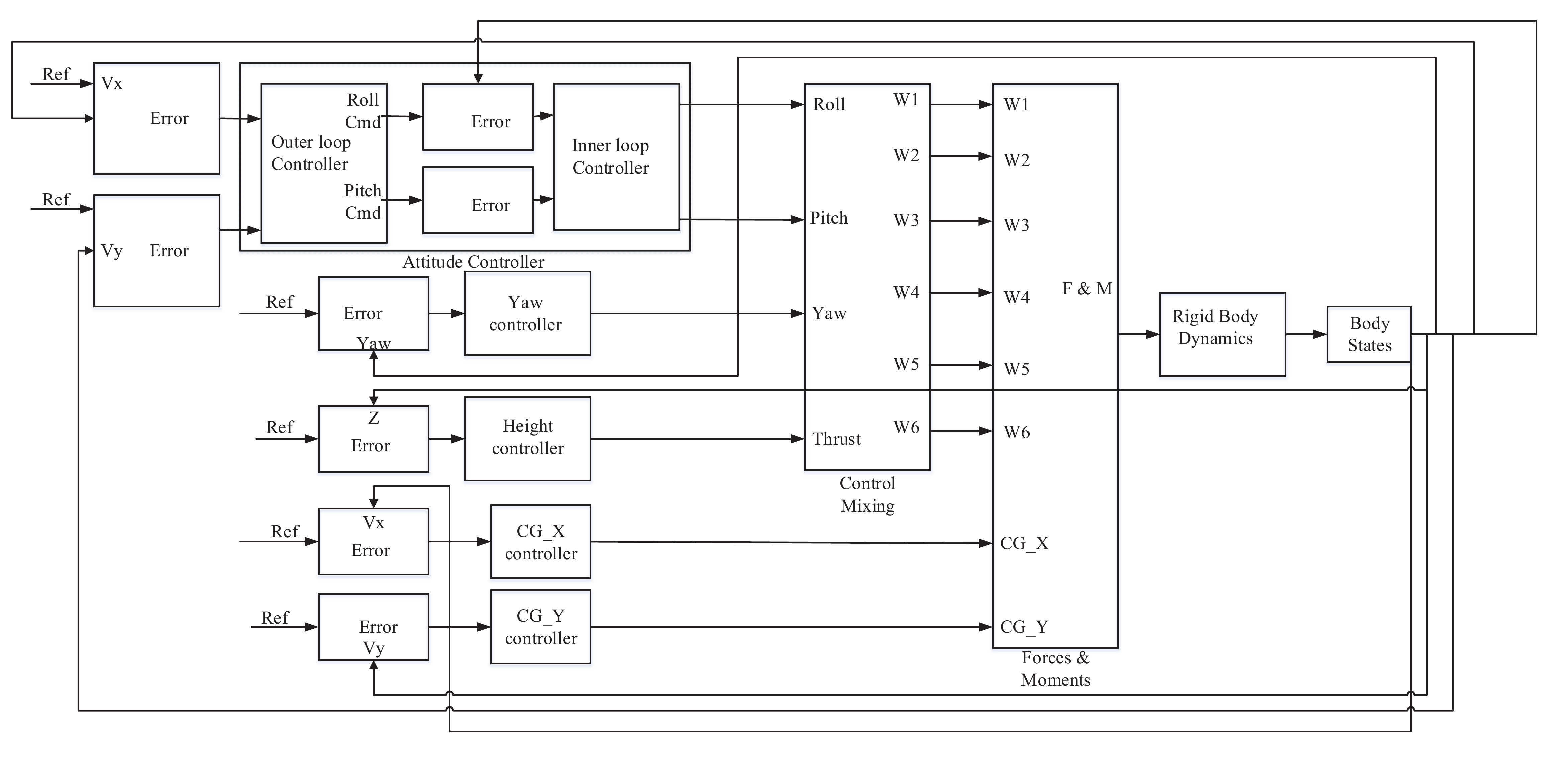}
\par\end{centering}

\caption{Top-level diagram of the over-actuated simulated Hexacopter plant\label{fig:Top-level-diagram-hexacopter} }

\end{figure*}

\par\end{center}

The 'attitude controller' block consists of inner loop and outer loop
controllers. In outer loop, our proposed G-controller is utilized
and PID is used in inner loop. The altitude or height, the moving-mass
based shifting of center of gravity in X and Y axis $(CG_{X},\;\text{and}\; CG_{Y})$
is controlled by the G-controller. Therefore, G-controllers are utilized
in 'outer loop controller', 'Height controller', '$CG_{X}$ controller',
'$CG_{Y}$ controller' blocks in Fig. \ref{fig:Top-level-diagram-hexacopter}.
The 'control mixing' block converts the attitude (roll and pitch),
yaw, and thrust commands coming from the controller to motor speed
commands. This is done using a simple linear mixing arrangement based
on the relevant positions of each rotor. The 'forces and moments'
block calculates the thrust and torque of each rotor based on the
relative airflow acting on each rotor and the commanded motor speed.
The thrusts and torque of each rotor are then added together to give
the total vertical force $(Fz)$ and yawing torque $(N)$ of the hexacopter.
The thrust of each rotor is multiplied by the appropriate moment arms
to calculate the rolling $(L)$ and pitching moments $(M)$ acting
on the airframe. The thrust, yawing torque and rolling and pitching
moments are then fed to the rigid body dynamics block so that the
hexacopter state can be updated. The above explained plant is also
being converted into hardware at the UAV laboratory of UNSW@ADFA.
The airframe components and basic structure of the hexacopter is exhibited
in Fig. \ref{fig:Airframe-components-hexa}. After the accomplishment
of the construction, our proposed controller will be instrumented
in that hardware. 

\begin{center}
\begin{figure}[tbh]
\begin{centering}
\includegraphics[scale=0.39]{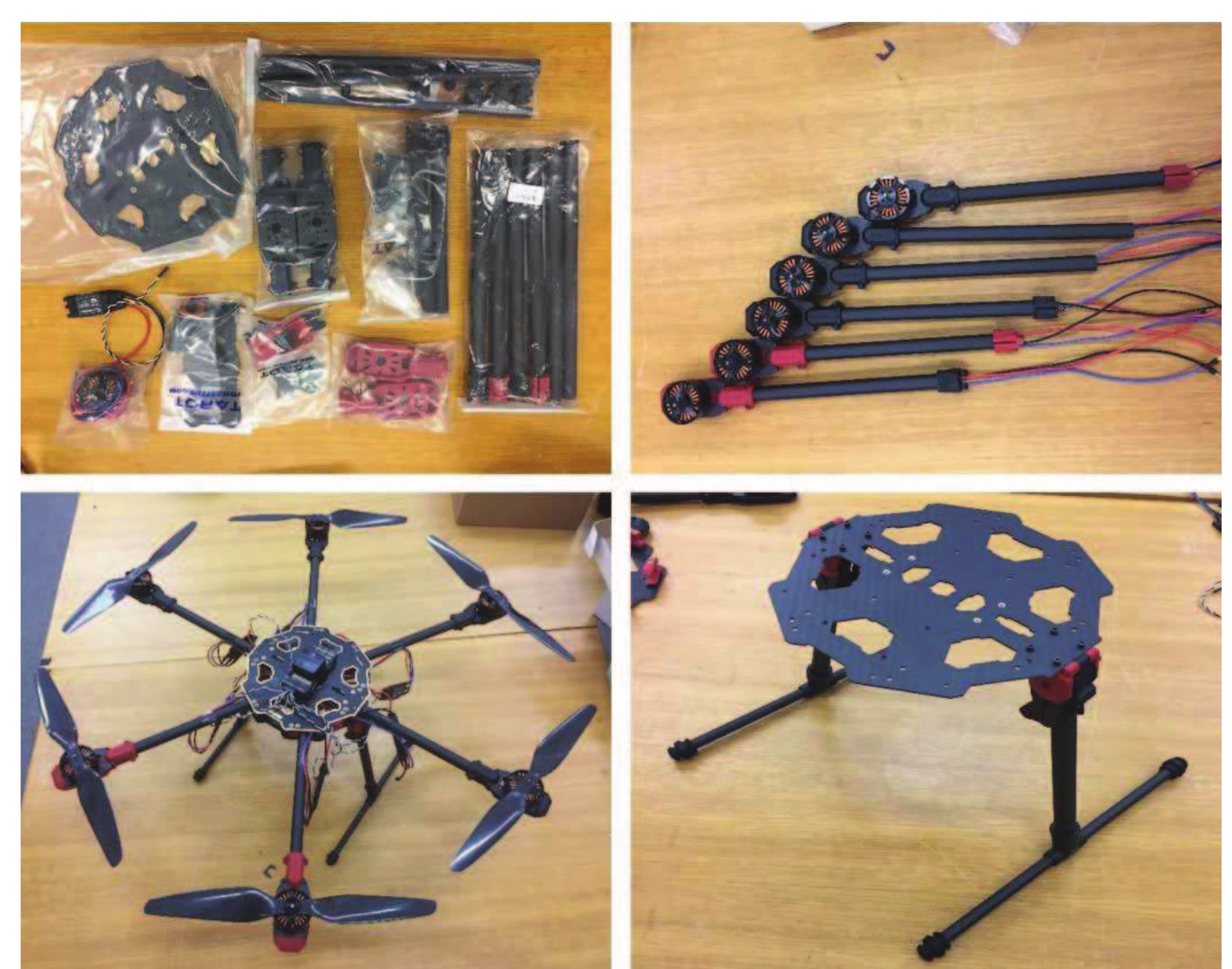}
\par\end{centering}

\caption{Airframe components and basic structure of the Hexacopter developed
at UAV Laboratory of UNSW@ADFA\label{fig:Airframe-components-hexa}}

\end{figure}

\par\end{center}

Deriving a proper mathematical model of such highly nonlinear, complex,
and over-actuated micro air vehicles are extremely difficult, where
insertion of uncertainties and unknown disturbance is more challenging,
or impossible in some cases. Thus, a controller that performs accurately
with a minimum or no knowledge about the system is much needed in
dealing with such autonomous vehicles. Being model-free and self-evolving
data-driven, our developed G-controller is an appropriate candidate.
Besides, the computational cost of our proposed controller is very
low, where it maintains better tracking performance. The self-evolving
architecture of our proposed G-controller is elaborated in the next
section.

\section{Architecture of the Self-Evolving G-Controller\label{sec:Architecture-of-Self-Evolving}}

The self organizing mechanism of the G-controller is adopted from
GENEFIS developed in \cite{pratama2014genefis}. GENEFIS is a TS FLS
that features multidimensional membership functions in the input space
where the contours are ellipsoid in arbitrary positions. Each estimated
one dimensional membership functions represent a portion in the input
space partition by assigning the Gaussian function's own center and
width. Concurrently, our GENEFIS based G-controller is generating
1st order first-order polynomials as consequent parts of the fuzzy
rules. In G-controller, a typical fuzzy rule can be expressed as follows:

\begin{equation}
\text{IF}~Z~\text{is}~R_{i},~\text{then}~\eta_{i}=a_{0i}+a_{1i}\zeta_{1}+a_{2i}\zeta_{2}+...+a_{ki}\zeta_{k}\label{eq:3}
\end{equation}
where $R_{i}$ represents the $i-th$ rule (membership function) constructed
from a concatenation of fuzzy sets and epitomizing a multidimensional
kernel, $k$ denotes the dimension of input feature, $Z$ is an input
vector of interest, $a_{i}$ is the consequent parameter, $\zeta_{k}$
is the $k-th$ input feature. The predicted output of the self-evolving
model can be expressed as:

\begin{align}
\hat{\eta} & =\sum\limits _{i=1}^{j}\psi_{i}(\zeta)\eta_{i}(\zeta)=\frac{\sum\limits _{i=1}^{j}R_{i}\eta_{i}}{\sum\limits _{i=1}^{j}R_{i}}\nonumber \\
 & =\frac{\sum_{i=1}^{j}\text{exp}(-(Z-\Theta_{i})\Sigma_{i}^{-1}(Z-\Theta_{i})^{T})\eta_{i}}{\sum_{i=1}^{j}\text{exp}(-(Z-\Theta_{i})\Sigma_{i}^{-1}(Z-\Theta_{i})^{T})}\label{eq:y1}
\end{align}
In Eq. \ref{eq:y1}, $\Theta_{i}$ is the centroid of the $i-th$
fuzzy rule $\Theta_{i}\in\Re^{1\times j}$, $\Sigma_{i}$ is a non-diagonal
covariance matrix $\Sigma_{i}\in\Re^{k\times k}$ whose diagonal components
are expressing the spread of the multivariate Gaussian function, and
$k$ is the number of fuzzy rules.

\subsection{Statistical Contribution Based Rule Growing Mechanism}

The Datum Significance (DS) method developed in \cite{huang2004efficient}
is utilized as a rule growing mechanism in G-controller. In our work,
the original DS method is geared into the multivariate Gaussian membership
function and polynomial consequents, which is the crux of GENEFIS
\cite{pratama2014genefis}. The integration of the multivariate Gaussian
membership function into the original DS method can be observed in
our work as follows:

\begin{equation}
D_{sgn}=|e_{rn}|\int_{Z}\exp\left(-\frac{(Z-Z_{n})\Sigma^{-1}(Z-Z_{n})^{T}}{(Z-\Theta)\Sigma^{-1}(Z-\Theta)^{T}}\right)\frac{1}{\mathcal{H}(Z)}dz\label{eq:DS1}
\end{equation}
where $D_{sgn}$ denotes the the significance of the $n-th$ datum,
$Z_{n}$ is the current input to the controller in a closed-loop control
system $Z_{n}\in\Re$, and $\mathcal{H}(Z)$ is the range of input
$Z.$ In a closed-loop control system, error $(e)$ indicates the
difference the desired reference and plant's output, which is usually
fed as input to the controller. Whereas, error $e_{rn}$ mentioned
in Eq. \ref{eq:DS1} is difference from the input error $(e)$. The
error $e_{rn}$ can be expressed as:

\begin{equation}
|e_{rn}|=|tr_{n}-\eta_{n}|\label{eq:e_rn}
\end{equation}
where $tr_{n}$ is the target, which is plant's output in our proposed
G-controller, and $\eta_{n}$ is control output from the G-controller
at $n-th$ episode. After applying $k-$fold numerical integration
to Eq. \ref{eq:DS1}, the following is obtained:

\begin{equation}
D_{sgn}=|e_{rn}|\left(\frac{\det(\Sigma_{j+1})}{\mathcal{H}(Z)}\right)^{k}
\end{equation}

When the statistical contribution of the datum is higher the existing
rules it becomes an appropriate candidate to be a new rule. Therefore,
the DS criterion can be amended mathematically as follows: 
\begin{equation}
D_{sgn}=|e_{rn}|\frac{\text{det}(\Sigma_{j+1})^{k}}{\sum_{i=1}^{j+1}\text{det}(\Sigma_{i})^{k}}\label{eq:DS2}
\end{equation}

When a sample lies far away from the nearest rule, a high value of
$D_{sgn}$ may obtain from Eq. \ref{eq:DS2} even with a small value
of $e_{rn}$. In such situation, generalisation capability of the
self-evolving neuro-fuzzy controller remains good without the addition
of any new rules. Therefore, a high value of $D_{sgn}$ does not always
indicate the necessity of a rule evolution. On the other hand, a high
value of $e_{rn}$ may obtain in case of an overfitting phenomenon.
In such case, the addition of a new rule may worsen the overfitting
phenomenon. Thus, a separation is needed in Eq. \ref{eq:DS2} to cover
two above mentioned discernible situation.

To overcome an overfitting scenario, it is important to monitor the
effect of a newly injected sample on $e_{rn}$, since the structural
learning is not occurring in every observation. In other words, the
rule growing mechanism is probably turned on when the rate of change
of $e_{rn}$ is positive. In this work, the mean and variance of $e_{rn}$
is measured by recursively updating $e_{rn}$ and standard deviation
\cite{angelov2011fuzzily} as follows: 

\begin{equation}
\bar{e}_{rn}=\frac{n-1}{n}\bar{e}_{rn-1}+\frac{1}{k}\bar{e}_{rn}
\end{equation}

\begin{equation}
\bar{\sigma}_{rn}^{2}=\frac{n-1}{n}\bar{\sigma}_{rn-1}^{2}+\frac{1}{k}(\bar{e}_{rn}-\bar{e}_{rn-1})
\end{equation}
When $\bar{e}_{rn}+\bar{\sigma}_{rn}^{2}-(\bar{e}_{rn-1}+\bar{\sigma}_{rn-1}^{2})>0,$
the DS criterion is simplified in our work as follows:

\begin{equation}
D_{sgn}=\frac{\text{det}(\Sigma_{j+1})^{k}}{\sum_{i=1}^{j+1}\text{det}(\Sigma_{i})^{k}}\label{eq:DS3}
\end{equation}
The condition in expanding the rule base utilizing Eq. \ref{eq:DS3}
is $D_{sgn}\ge g$, where $g$ is a predefined threshold. Eq. \ref{eq:DS3}
represents an encouraging generalization and summarization of the
datum, since a new rule can omit possible overfitting effects. Besides,
this DS criterion can predict the probable contribution of the datum
during its lifetime.

\subsection{Statistical Contribution Based Rule Pruning Mechanism}

The Extended Rule significance (ERS) method was put forward by \cite{huang2004efficient}.
The ERS concept appraises the statistical contribution of the fuzzy
rules when the number of observations is approaching infinity. The
default ERS theory is not possible to integrate directly into GENEFIS's
learning platform due to the incompatibility of default ERS concept
with the NN. Numerous modifications are made in ERS theory to fit
them with GENEFIS based G-controller. In this work, the concept of
statistical contribution of the fuzzy rules can be expressed mathematically
as follows:

\begin{equation}
\mathcal{E}(i,n)=|\delta_{i}|\mathcal{E}_{i},\;\text{where}\;|\delta_{i}|=\sum_{i=1}^{k+1}|\eta_{i}|\label{eq:E_1}
\end{equation}

\begin{equation}
\mathcal{E}_{i}=\int_{Z}\exp(-(Z-\Theta_{i}^{n})\Sigma_{i}^{-1}(Z-\Theta_{i}^{n})^{T})\frac{1}{\mathcal{H}(Z)}dz\label{eq:E_2}
\end{equation}
From Eq. \ref{eq:E_1}, it can be anticipated that the contribution
of the fuzzy rules is a summary of the total contribution of input
and output parts of the fuzzy rules, where $\mathcal{E}_{i}$ is expressing
the modified version of original input contribution explained in \cite{huang2004efficient,huang2005generalized},
and $\delta_{i}$ is explicating the contribution of output parameters.
Usually, inverse covariance matrix $\Sigma_{i}^{-1}$ in Eq. \ref{eq:E_2}
has a smaller size than that of $Z$, which necessitates an amendment
in Eq. \ref{eq:E_2} as follows:

\begin{equation}
\mathcal{E}_{i}\approx\frac{1}{\mathcal{H}(Z)}\left(2\int_{0}^{\infty}\exp\left(-\left(\frac{Z^{2}}{\det(\Sigma_{i})}\right)\right)dz\right)^{k}
\end{equation}

By using the $k-$fold numerical integration, the final version of
ERS theory can be expressed as: 
\begin{equation}
\mathcal{E}_{inf}^{i}=\sum_{i=1}^{j+1}\eta_{i}\frac{\text{det}(\Sigma_{i})^{k}}{\sum_{i=1}^{j}\text{det}(\Sigma_{i})^{k}}\label{eq:E_3}
\end{equation}

When $\mathcal{E}_{inf}^{i}\le k_{e}$, it is presumed that the clusters
cannot capture the latest incoming data to the G-controller in the
closed-loop control cycle. It can be deduced that the hypervolume
of the triggered cluster indicates the significance of the fuzzy rule.
Thus, when the volume of the $i$th cluster is much lower than the
summation of volumes of all cluster, that rule is considered as inconsequential.
Such a rule is pruned to protect the rule base evolution from its
adverse effect. In this work, $k_{e}$ exhibits a plausible trade-off
between compactness and generalization of the rule base. The allocated
value for $\delta$ is $\delta=[0.0001,1]$, and $k_{e}=10\%\;\text{of}\;\delta$.

\subsection{Adaptation of Rule Premise Parameters}

Generalized Adaptive Resonance Theory+ (GART+) \cite{oentaryo2011bayesian}
is used in G-controller as a technique of granulating input features
and adapting premise parameters. It is observed that GART \cite{yap2008hybrid}
and its successor improved GART (IGART) \cite{yap2011improved} suffer
from a category growing problem. In GRAT the compatibility measure
is done utilizing the maximal membership degree of a new datum to
all available rules. In first round if the selected category expresses
a higher membership degree than a predefined threshold $\rho_{a}$,
then it is declared as winning category and the match tracking mechanism
is executed. However, the first round winning category fails to beat
the match tracking threshold $\rho_{b}$, which deactivates that category
and increases the value of threshold $\rho_{a}$ to find a better
candidate. A larger width is required in the next selected category
to cope with the increased value of $\rho_{a}$. Otherwise, it fabricates
a new category. Nonetheless, a category with larger radii may contain
more than one distinguishable data clouds and thereby marginalizing
the other clusters in every training episode. In incremental learning
environment this effect is known as $\textit{cluster delamination}$
\cite{lughofer2012dynamic}, pictorially exhibited in Fig. \ref{fig:cluster delamination}.
To relieve from the $\textit{cluster delamination}$ effect, in GENEFIS
based G-controller the size of fuzzy rule are constrained by using
GART+, which allows a limited grow or shrink of a category.

\begin{center}
\begin{figure}[b]
\begin{centering}
\textsf{\includegraphics[scale=0.27]{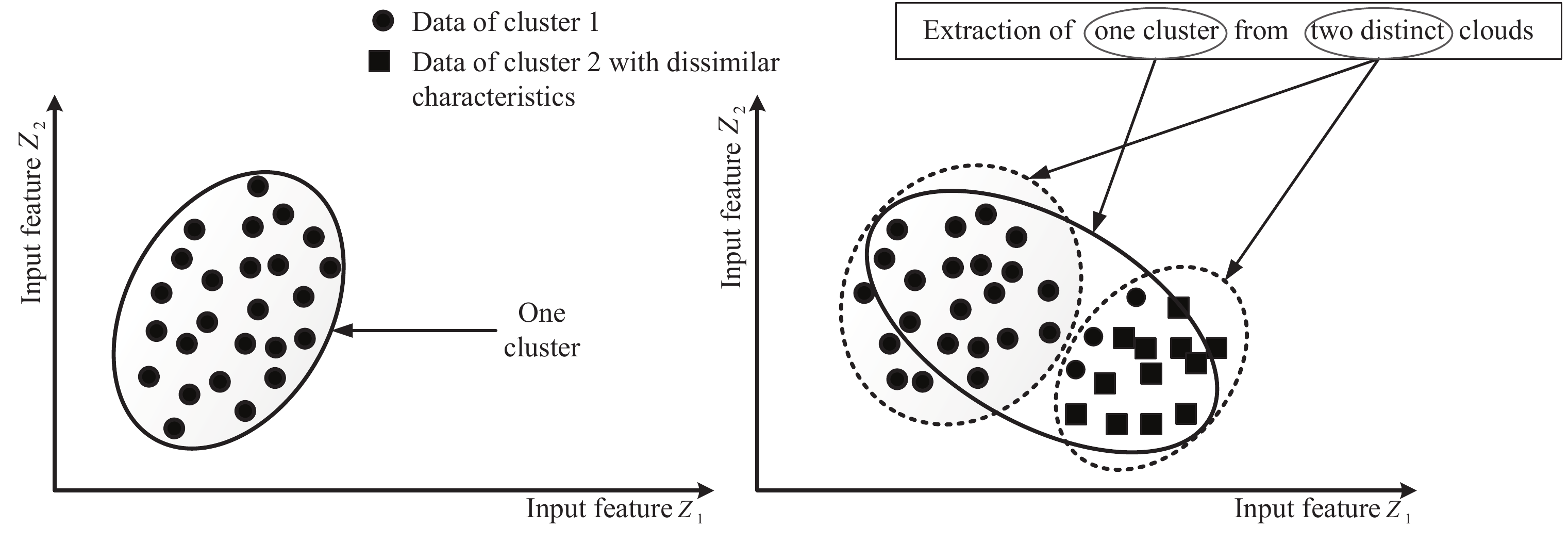}}
\par\end{centering}

\caption{Cluster delamination effect (adapted from \cite{pratama2014genefis}
with proper permission)\label{fig:cluster delamination}}
\end{figure}

\par\end{center}

\subsubsection{Improved Selection procedure of winning Category}

To determine the most compatible or winning category Bayes's decision
theory is utilized in GART+ \cite{oentaryo2011bayesian}. To stimulate
a more appropriate selection of the winning category, the Bayesian
concept does not only consider the proximity of a category to the
inserted datum, but also the dominance of the category with respect
to the other categories through the category prior probability. That
is, the category prior probability can count the number of samples
falling in the outreach of the category, and is expressed as follows:

\begin{equation}
\hat{P}_{r}(\psi_{i})=\frac{N_{i}}{\sum_{i=1}^{j}N_{i}}\label{eq:prior}
\end{equation}
where $N_{i}$ indicates the number of times that ith category or
fuzzy rule wins the competition.

When a new input datum to the controller finds two categories with
almost similar distance but with different population number, Bayes\textquoteright{}s
decision theory assists to select the category with more data points
and declared it as a winning category. In the proposed G-controller,
posterior probability of the $i$th category can be represented as
follows:

\begin{equation}
\hat{P}_{r}(\psi_{i}|Z)=\frac{\hat{p}_{r}(Z|\psi_{i})\hat{P}_{r}(\psi_{i})}{\sum_{i=1}^{j}\hat{p}_{r}(Z|\psi_{i})\hat{P}_{r}(\psi_{i})}\label{eq:posterior}
\end{equation}
where $\hat{p}_{r}(\psi_{i}|Z)$ and $\hat{P}_{r}(\psi_{i})$ represents
the likelihood and the prior probability correspondingly. The likelihood
can also be elaborated as follows:

\begin{equation}
\hat{p}_{r}(Z|\psi_{i})=\frac{1}{(2\pi V_{i})^{1/2}}\text{exp}(-(Z-\Theta_{i})\Sigma_{i}^{-1}(Z-\Theta_{i})^{T})\label{eq:likelihood}
\end{equation}
where $V_{i}$ determines the estimated hyper-volume of feature space
covered by the $i$th category, which can be expressed as: 
\begin{equation}
V_{i}=\text{det}(\Sigma_{i})\label{eq:volume}
\end{equation}

The Bayesian concept presented in Eq. \ref{eq:posterior} is implemented
in G-controller, which can be interpreted as follows:
\begin{enumerate}
\item When a new sample is adjacent to existing categories, it causes a
higher likelihood expressed by Eq. \ref{eq:likelihood}.
\item A category with a large volume is forced to divide its volume in Eq.
\ref{eq:likelihood}, as a consequent it delivers a lower value of
the posterior probability \ref{eq:posterior}. This is important particularly
to avoid large span clusters and to decrease the likelihood of cluster
delamination effects. 
\item According to \ref{eq:prior}, categories surrounded by more incoming
data samples are more worthwhile, which inflicts a high value of posterior
probability.
\end{enumerate}

\subsubsection{Vigilance Test}

There are two goals to perform the vigilance test. The first goal
concerns about the capability of the winning category to accommodate
a new datum. The second goal is to reduce the size of the category,
where a rule is not allowed to have a volume higher than the threshold
$V_{max}$, that is calculated from $V_{max}\equiv\rho_{b}\sum_{i=1}^{j}V_{i}$.
The vigilance test is a way to rule deletion, update, or evolution
where it needs to be satisfied four different conditions as presented
below:

\begin{equation}
\text{Case I:}\; R_{win}\ge\rho_{a},\; V_{win}\le V_{max}\label{eq:case1}
\end{equation}
where $R_{win}$ is the membership degree of the winning rule to seize
the latest datum. More importantly, the condition in Eq. \ref{eq:case1}
is indicating the capability of the selected category to accommodate
the newest datum and emphasizing on the limited size of a category.
In our proposed G-controller $\rho_{a}$ is set close to 1. Contrarily,
the value of $\rho_{b}$ is set as {[}0.0001, 0.1{]}. Then the adaptation
mechanism of focal point $\Theta_{i}$, and the dispersion matrix
$\Sigma_{i}$ is generated by the equations as follows:

\begin{align}
 & \Theta_{win}^{new}=\frac{N_{win}^{old}}{N_{win}^{old}+1}\Theta_{win}^{old}+\frac{\left(Z-\Theta_{win}^{old}\right)}{N_{win}^{old}+1}\label{eq:C-new}\\
 & \Sigma_{win}^{new{}^{-1}}=\frac{\Sigma_{win}^{old^{-1}}}{1-\alpha}+\frac{\alpha}{1-\alpha}\nonumber \\
 & \frac{\left(\Sigma_{win}^{old^{-1}}\left(Z-\Theta_{win}^{new}\right)\right)\left(\Sigma_{win}^{old^{-1}}\left(Z-\Theta_{win}^{new}\right)\right)^{T}}{1+\alpha\left(Z-\Theta_{win}^{new}\right)\Sigma_{win}^{old^{-1}}\left(Z-\Theta_{win}^{new}\right)^{T}}\label{eq:sigma_new}\\
 & N_{win}^{new}=N_{win}^{old}+1\label{eq:population_new}
\end{align}
where $\alpha$ can be expressed as follows:

\begin{equation}
\alpha=\frac{1}{N_{win}^{old}+1}
\end{equation}
where $N_{win}^{old}$ denotes the number of training samples populating
the winning cluster.

Besides, a major advantage of utilizing Eq. \ref{eq:sigma_new} is
the prompter update of the dispersion matrix (inverse covariance matrix),
since a direct adjustment of the dispersion matrix is occurring without
the necessity to re-inverse the dispersion matrix \cite{lughofer2011evolving}.
Concerning the conditions in \ref{eq:case1}, some pertinent likelihoods
may emerge in the rehearsal process and they are outlined as follows:

$\text{Case II:}\; R_{win}<\rho_{a},\; V_{win}>V_{max}$

In this circumstance, the input data to the G-controller cannot be
touched by any existing rules of the controller, since the inserted
input data is hardly covered by any rules. The statistical contribution
of the datum needs to be calculated by DS-criterion. When both conditions
are satisfied, a new rule is generated and its parameters are assigned
as follows:

\begin{align}
\Theta_{j+1} & =Z\label{eq:C_j+1}\\
\text{diag}\bigg(\Sigma_{j+1}\bigg) & =\frac{\text{max}((\Theta_{i}-\Theta_{i-1}),(\Theta_{i}-\Theta_{i+1}))}{\sqrt{\frac{1}{\text{In}(\epsilon)}}}\label{eq:sigma-j+1}
\end{align}
where the value of $\epsilon$ is 0.5. Equation \ref{eq:sigma-j+1}
ensures a sufficient coverage of the newly added rule, which is proved
in \cite{wu2001fast}. It helps GENEFIS to explore untouched regions
in the feature space fitting a superfluous cluster at whatever point
a relatively unexploited region or knowledge is fed, which is a mandatory
element to confronting possible non-stationary and evolving qualities
of the self-evolving control system. Note that proper initialization
of inverse covariance matrix plays a crucial role to the success of
multivariate Gaussian fuzzy rule. Although it meets the $\epsilon-$completeness
criterion, Equation \ref{eq:sigma-j+1} requires re-inversion phase
which sometimes leads to instability when the covariance matrix is
not full-rank. As an alternative, the inverse covariance matrix can
be initialized as:

\begin{equation}
\Sigma_{0}^{-1}=k_{fs}I
\end{equation}
where $k_{fs}$ is a user-defined parameter, and $I$ is an identity
matrix.

$\text{Case III:}\; R_{win}\ge\rho_{a},\; V_{win}>V_{max}$

This situation is indicating the capability of existing rule base
to cover the current data easily. However, the width of the chosen
cluster is oversized. This datum creates a redundancy when added to
the rule base. To mitigate the adverse impact, one of the solutions
is to replace the selected cluster merely by this datum. Then the
fuzzy region is eased as follows: 
\begin{align}
 & \Theta_{win}=Z\label{eq:18}\\
 & \Sigma_{win}^{new{}^{-1}}=\frac{1}{k_{win}}\Sigma_{win}^{old^{-1}}\label{eq:19}
\end{align}
 where $k_{win}$ is a constant with a value of 1.1, and the width
of the cluster is reduced until a desirable fuzzy region is obtained
while satisfying $V_{win}\le V_{max}$.

$\text{Case IV:}\; R_{win}<\rho_{a},\; V_{win}\le V_{max}$

The same action is taken as in Case I, i.e., the adjustment process
is executed to stimulate the category to move towards the incoming
input data.

\section{SMC Theory-Based Adaptation of G-Controller\label{sec:SMC-Theory-based-Adaptation} }

In our work, an advanced self-evolving neuro-fuzzy system called GENEFIS
is utilized to build the evolving structure and adapt premise parameters
of the proposed G-controller, where the integration of multivariate
Gaussian function and GART+ method helps the controller to reduce
the structural complexity and to adapt with sharp changes in autonomous
vehicle's plant dynamics. On the other hand, being robust enough to
guarantee the robustness of a system against external perturbations,
parameter variations, unknown uncertainties, the SMC theory is applied
in our work to adapt the consequent parameters of the G-controller.
In SMC scheme, the motion of a system is restricted to a plane known
as $\textit{sliding surface}$. In this work, the SMC learning theory-based
adaptation laws are developed to establish a stable closed-loop system.
By following the regulations of SMC scheme as explained in \cite{kayacan2012sliding,utkin2013sliding,kayacan2017type},
the zero dynamics of the learning error coordinate is defined as time-varying
sliding surface as follows:

\begin{equation}
S_{ssr}(u_{g},u)=u_{ARC}(t)=u_{g}(t)+u(t)
\end{equation}

The sliding surface for the highly nonlinear over-actuated autonomous
vehicles namely FW MAV, and hexacopter plant to be controlled is expressed
as:

\begin{equation}
s_{H}=e+\lambda_{1}\dot{e}+\lambda_{2}\int_{0}^{t}e(\tau)d\tau
\end{equation}
where, $\lambda_{1}=\frac{\alpha_{2}}{\alpha_{1}},$ $\lambda_{2}=\frac{\alpha_{3}}{\alpha_{1}}$$,$$\: e$
is the error which is the difference between the actual displacement
from the plant and desired position in case of altitude control. In
this work, in case the BIFW MAV plant, the sliding parameter $\alpha_{1}$
has initialized with a small value $1\times10^{-2}$, whereas $\alpha_{2}$
has initialized with $1\times10^{-3},$ and $\alpha_{3}\thickapprox0.$
Each of the parameters is then evolved by using learning rates. These
learning rates are set in such a way so that the sliding parameters
can achieve the desired value in the shortest possible time to create
a stable closed-loop control system. A higher initial value of the
sliding parameters is avoided, since it may cause a big overshoot
at the beginning of the trajectory. It can be abstracted that, to
make our proposed G-controller absolutely model free, these sliding
parameters are self-organizing rather than predefined constant values.

$Definition:$ After a certain time $t_{k}$ a sliding motion will
be developed on the sliding manifold $S_{ssr}(u_{g},u)=u_{ARC}(t)=0$,
where the state $S_{ssr}(t)\dot{S}_{ssr}(t)=u_{ARC}(t)\dot{u}_{ARC}(t)<0$
to be satisfied for the whole time period with some nontrival semi-open
sub-interval of time expressed as $[t,\; t_{k})\subset(0,\; t_{k})$.

It is expected to produce such online adaptation of consequent parameters
of the proposed G-controller that the sliding mode condition of the
aforestated definition is enforced. The adaptation process of the
proposed method is summarized below.

\begin{center}
\begin{figure}[t]
\begin{centering}
\textsf{\includegraphics[scale=0.39]{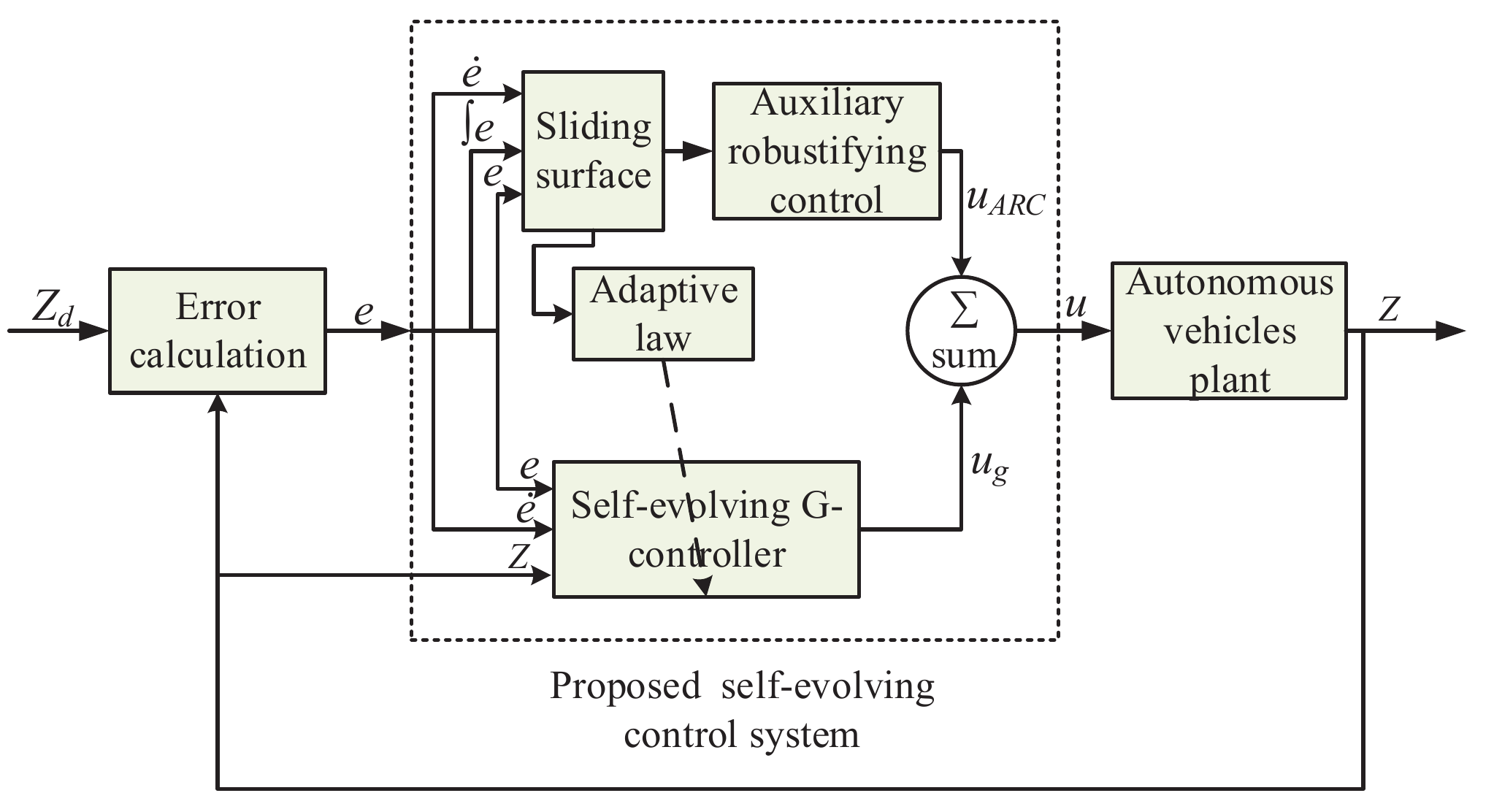}}
\par\end{centering}

\caption{Self-evolving G-controller based closed-loop control system\label{fig:g-controller-closed-loop}}
\end{figure}

\par\end{center}
\begin{thm}
The adaptation laws for the consequent parameters of the G-controller
are chosen as:
\end{thm}
\begin{equation}
\dot{\omega}(t)=-\alpha_{1}G(t)\psi(t)s_{H}(t),\;\;\text{where }\;\omega(0)=\omega_{0}\in\Re^{nR\times1}
\end{equation}
where the term $G(t)$ can be updated recursively as follows: 

\begin{equation}
\dot{G}(t)=-G(t)\psi(t)\psi^{T}(t)G(t),\;\;\text{where }\; G(0)=G_{0}\in\Re^{nR\times nR}
\end{equation}
where $n$ is the number of inputs to the controller, and $R$ is
the number of generated rules. These adaptation laws guarantee a stable
closed-loop control system, where the plants to be controlled can
be of various order. 

$Proof:$ The sliding parameter dependent robustifying auxiliary control
term of the proposed controller can be expressed as follows:

\begin{equation}
u_{ARC}(t)=\alpha_{1}s_{H}\label{eq:u_aux}
\end{equation}

The robustifying auxiliary control term $u_{ARC}$ may suffer from
high-frequency oscillations in the control input. It is an undesirable
phenomenon in sliding mode controller and known as chattering effect.
Due to simplicity, saturation or sigmoid functions are mostly used
to reduce the chattering effect. In this work, a saturation function
is utilized to mitigate the adverse effect of chattering. 

The G-controller's final output signal can be expressed as follows:

\begin{equation}
u_{g}(t)=\psi^{T}(t)\omega(t)\label{eq:u_g}
\end{equation}

The overall control signal as observed in Fig.\eqref{fig:g-controller-closed-loop}
can be obtained as follows:

\begin{equation}
u(t)=u_{ARC}(t)-u_{g}(t)\label{eq:u}
\end{equation}

The cost function can be defined as:

\begin{align}
J(t) & =\int_{0}^{t}s_{H}^{2}(\tau)d\tau\nonumber \\
 & =\frac{1}{\alpha_{1}^{2}}\int_{0}^{t}(u(\tau)+u_{g}(\tau))^{2}d\tau\nonumber \\
 & =\frac{1}{\alpha_{1}^{2}}\int_{0}^{t}(u(\tau)+\psi^{T}(t)\omega(\tau))^{2}d\tau
\end{align}

The gradient of J with respect to $\omega$ is as follows:

\begin{align}
 & \nabla_{\omega}J(t)=0\nonumber \\
 & \Rightarrow\int\psi(\tau)u(\tau)d\tau+\omega(t)\int_{0}^{t}\psi(\tau)\psi^{T}(\tau)d\tau=0\nonumber \\
 & \Rightarrow\omega(t)=\left[\int_{0}^{t}\psi(\tau)\psi^{T}(\tau)d\tau\right]^{-1}\int_{0}^{t}\psi(\tau)u(\tau)d\tau\\
 & \Rightarrow\omega(t)=-G(t)\int_{0}^{t}\psi(\tau)u(\tau)d\tau\label{eq:psi(t)}\\
 & \Rightarrow G^{-1}(t)\omega(t)=-\int_{0}^{t}\psi(\tau)u(\tau)d\tau\label{eq:Gt^-1_psi}
\end{align}
where,

\begin{equation}
\begin{aligned}G(t)\end{aligned}
=\left[\int_{0}^{t}\psi(\tau)\psi^{T}(\tau)d\tau\right]^{-1}
\end{equation}

\begin{equation}
\begin{aligned}G^{-1}(t)\end{aligned}
=\int_{0}^{t}\psi(\tau)\psi^{T}(\tau)d\tau\label{eq:G(t)^-1}
\end{equation}

The derivative of Eq. \eqref{eq:G(t)^-1} is as follows:

\begin{align}
G^{-1}(t)\dot{G}(t)G^{-1}(t) & =-\psi(t)\psi^{T}(t)\nonumber \\
\dot{G}(t) & =-G(t)\psi(t)\psi^{T}(t)G(t)\label{eq:G_dot(t)}
\end{align}
From Eq. \eqref{eq:G_dot(t)}, it is observed that $\dot{G}(t)$ is
a negative definite and G(t) is decreasing over time, therefore $G(t)\in l_{\infty}$.
Now executing the time derivative of Eq. \eqref{eq:psi(t)} and utilizing
Eq. Fig. 6(a)\eqref{eq:u_aux}, \eqref{eq:u_g}, \eqref{eq:u}, and
\eqref{eq:Gt^-1_psi} the following is obtained:

\begin{align}
\dot{\omega}(t) & =\dot{G}(t)G^{-1}(t)\omega(t)-G(t)\psi(t)u(t)\nonumber \\
 & =-G(t)\psi(t)\psi^{T}(t)\psi(t)-G(t)\psi(t)u(t)\nonumber \\
 & =-G(t)\psi(t)\left(\psi^{T}(t)\omega(t)+u(t)\right)\nonumber \\
 & =-\alpha_{1}G(t)\psi(t)s_{H}(t)
\end{align}

\subsubsection{Stability Analysis}

$Definition:$ FLS is known as a general function approximator. Therefore,
in this work it is assumed that without loss of generality there exists
a $\omega^{*}$ such that:

\begin{equation}
u(t)=\psi^{T}\omega^{*}(t)+\varepsilon_{f}^{*}(z)
\end{equation}
where $\varepsilon_{f}^{*}(z)=[\varepsilon_{f1}^{*},\varepsilon_{f1}^{*},...,\varepsilon_{f1}^{*}]^{T}\in\Re^{k}$
is the minimal functional approximator error. In this work, the following
is defined:

\begin{equation}
\tilde{\omega}(t)=\omega(t)-\omega^{*}
\end{equation}

In addition:

\begin{equation}
s_{H}(t)=\psi^{T}\widetilde{\omega}(t)
\end{equation}

$Lemma\;1:$
\begin{align}
\frac{d(G^{-1}(t)\tilde{\omega}(t))}{dt} & =-G^{-1}(t)\dot{G}(t)G^{-1}(t)\tilde{\omega}(t)+G^{-1}(t)\dot{\widetilde{\omega}}(t)\nonumber \\
 & =\psi(t)\psi^{T}(t)\tilde{\omega}(t)-\psi(t)s_{H}(t)\nonumber \\
 & =\psi(t)s_{H}(t)-\psi(t)s_{H}(t)\nonumber \\
 & =0
\end{align}

This is indicating that $G^{-1}(t)\tilde{\omega}(t)$ is not altering
with respect to time, and therefore $G^{-1}(t)\tilde{\omega}(t)=G^{-1}(0)\tilde{\omega}(0),$
$\forall_{t}>0.$

\begin{equation}
lim_{t\rightarrow\infty}\tilde{\omega}(t)=lim_{t\rightarrow\infty}G(t)G^{-1}(0)\tilde{\omega}(0)
\end{equation}

Since $G(t)$ is decreasing and$\tilde{\omega}(t)\in l_{\infty},$$\omega(t)\in l_{\infty}.$
In this work the following Lyapunov function is considered:

\begin{equation}
V(t)=\frac{1}{2}\widetilde{\omega}^{T}(t)G^{-1}(t)\tilde{\omega}(t)\label{eq:V}
\end{equation}

The time derivative of the Lyapunov function is as follows:

\begin{align}
\dot{V}(t) & =\frac{1}{2}\widetilde{\omega}^{T}(t)G^{-1}\dot{\tilde{\omega}}(t)+\frac{1}{2}\widetilde{\omega}^{T}(t)\dot{G}^{-1}\tilde{\omega}(t)\nonumber \\
 & =-\widetilde{\omega}^{T}(t)\psi(t)s_{H}(t)-\frac{1}{2}\widetilde{\omega}^{T}(t)\psi(t)\psi^{T}(t)\tilde{\omega}(t)\nonumber \\
 & =-s_{H}^{2}(t)-\frac{1}{2}s_{H}^{2}(t)\nonumber \\
 & =-\frac{3}{2}s_{H}^{2}(t)\leq0\label{eq:V-dot}
\end{align}

From Eq. \ref{eq:V}, and Eq. \ref{eq:V-dot}, it is observed that
$V(t)>0,$ and $\dot{V}(t)\leq0$. In addition, Eq. \ref{eq:V-dot}
shows that $\dot{V}(t)=0,$ if and only if $e(t)=0.$ It is indicating
that the global stability of the system is guaranteed by the Lyapunov
theorem. By utilizing Barbalat\textquoteright{}s lemma \cite{slotine1991applied},
it can also be observed that $e(t)\rightarrow0$ as $t\rightarrow\infty.$
It is ensuring the asymptotic stability of the system. Thus, a convergence
of the system's tracking error to zero is witnessed.

\section{Results and Discussion\label{sec:Results-and-Discussion}}

In this work, the self-evolving generic neuro-fuzzy controller namely
G-controller is attempted to control of autonomous vehicles online.
As mentioned in Section \ref{sec:Problem-Statement} that the G-controller
is utilized in controlling a BIFW MAV and a hexacopter UAV. In case
of the BIFW MAV, various altitude trajectory tracking is observed
to evaluate the controller's performance, whereas, in hexacopter,
not only the altitude but also the attitude tracking is witnessed.
Being an evolving controller, the G-controller can evolve both the
structure and parameters. The observed evaluation procedure from the
G-controller's performance is explained in the following subsection
\ref{sub:Evaluation-Procedure}.

\subsection{Evaluation Procedure\label{sub:Evaluation-Procedure}}

The proposed G-controller has the capability of evolving the structure
by adding or pruning the rules like many other evolving controllers
discussed in the subsection \ref{sub:Related-Work}. However, unlike
the existing evolving controller, GRAT+, multivariate Gaussian function,
SMC learning theory based adaptation laws are combined in the G-controller.
From the amalgamation of such advanced features, the faster self-evolving
mechanism is recorded with a lower computational cost. In controlling
the altitude and attitude of the highly nonlinear and complex autonomous
vehicles discussed in Section \ref{sec:Problem-Statement}, the activation
of only the rule growing mechanism was sufficient. Due to the evolving
nature of the G-controller, the fuzzy rules are generated in different
time steps for different reference signals. This rule generation of
the G-controller with respect to various desired altitude of BIFW
MAV are compiled in a table provided in the supplementary document.
To understand graphically, the number of generated rules for various
trajectories of BIFW MAV and Hexacopter are plotted and disclosed
in Fig. \ref{fig:Generated-rules}.

\subsection{Results}

The G-controller's performance is observed with respect to various
reference signals, and the results are compared with a TS fuzzy controller
\cite{ferdaus2017fuzzyclusteringFWMAV}, and a Proportional Integral
Derivative (PID) controller. Our source codes are made publicly available
in \cite{mpratamaweb}. In case of the BIFW MAV, variety of desired
trajectories were utilized in the closed-loop control system to evaluate
controllers performance for 100 seconds, such as: 1) a constant altitude
of 10 meters expressed as $Z_{d}(t)=10$; 2) three different step
functions, where one of them is varying its amplitude from 0 m to
10 m, another is from 5 m to 10 m, and the other one is varying from
-5 m to 5m, presented as $Z_{d}(t)=10u(t-20)$, $Z_{d}(t)=5u(t)+5u(t-20)$,
$Z_{d}(t)=-5u(t)+10u(t-20)$ respectively; 3) three different square
wave function with a frequency of 0.1 Hz, and amplitude of 1 m, 4
m, and 10 m correspondingly; 4) two square wave function with a frequency
of 1 Hz, and amplitude of 1 m and 4 m; 5) a customized trajectory,
where the amplitude varies from 0 to 2 m; 6) a sawtooth wave function
with an amplitude of 1 m and frequency of 1 Hz; 7) a sine wave function
with an amplitude of 1 m and frequency of 1 Hz. For all these trajectories,
the performance of our proposed G-controller, TS fuzzy controller,
and PID controller is observed and compared, where higher accuracy
is obtained from the G-controller. For clearer understanding, some
of these observations are presented pictorially from Fig. \ref{fig:step-fw}
to Fig. \ref{fig:gust}.

The performance of various controllers for a step function $Z_{d}(t)=5u(t)+5u(t-20)$
is observed in Fig. \ref{fig:step-fw}, where our proposed G-controller
outperformed the PID, and TS fuzzy controller. The performance of
the trajectory like square wave pulse with an amplitude of 4 m and
frequency of 1 Hz is observed in Fig. \ref{fig:square_f1_a4_FW},
where comparatively improved performance is witnessed from our developed
G-controller. In case of this trajectory the TS fuzzy controller fails
to follow the trajectory. Therefore, the comparison only between the
PID and G-controller is exposed in Fig. \ref{fig:square_f1_a4_FW},
where the G-controller has beaten the PID controller in terms of accuracy.
In case of all the trajectories for BIFW MAV exposed in this paragraph,
superior performance is visualized by our proposed G-controller. To
observe the controllers performance deeply, the root mean square error
(RMSE), rising time, and settling time of all the controllers for
various reference signals are measured and summarized in TABLE \ref{tab:rmse},
where the lowest RMSE is inspected from the G-controller. Since the
G-controller starts operating from scratch with an empty fuzzy set,
the rising time is comparatively higher than the PID controller. However,
comparatively lower settling time is indicating the proposed controllers
ability to back to the desired trajectory sharply.

A wind gust is added to the BIFW MAV plant dynamics to check the robustness
of our proposed G-controller against unknown perturbations and uncertainties.
This simulated wind gust has a maximum velocity of 40 $ms^{-1}$ and
is applied to the plant after 2 seconds. In the presence of the wind
gust, some of the trajectory tracking performances of the controllers
are manifested in Fig. \ref{fig:gust}, where an insignificant deterioration
in tracking is observed at the beginning. However, this adverse effect
has been minimized very sharply by the G-controller. The RMSE by considering
the effect of wind gust are also tabulated in TABLE \ref{tab:rmse}. 

Furthermore, the G-controller has been utilized to control altitude,
the outer loop of attitude (roll and pitch) of the simulated over-actuated
hexacopter plant developed in ADFA and results are compared with a
PID controller. In case of controlling the altitude, the controllers
are employed to control the thrust of the control-mixing box of the
plant. Due to the addition of the moving mass, the rolling motion
is not only controlled by the velocity in Y-axis $(v_{y})$ generated
by the motors, but also the mass moving in the Y direction due to
their Center of Gravity (CG) shifting capability. Our proposed controller
has been employed in both facts to control the rolling motion. Similarly,
to control the pitching motion of the hexacopter, the G-controller
has been used to control both the velocity in X-axis $(v_{x})$ and
the mass moving in the X direction. The altitude tracking performance
of hexacopter for various trajectories has been observed in Fig. \ref{fig:altitude_hexa},
whereas the tracking of rolling and pitching is exhibited in Fig.
\ref{fig:roll_pitch_hexa}. In all cases, better tracking has been
monitored from the G-controller than that of PID controller. The RMSE,
rise time, and settling time has been calculated for all the trajectories
of the hexacopter and outlined in a TABLE provided in the supplementary
document, where lower RMSE is perceived from the proposed G-controller.
Besides, the settling time of the G-controller is much lower than
that of the PIDs, which clearly indicates their improvement over the
PID controller.

\subsection{Discussion}

Unlike the PID and TS fuzzy controller, the G-controller starts the
self-construction online with an empty fuzzy set at the beginning
of the closed-loop control system. Whereas, the PID and TS fuzzy controller
start operating with their pre-set control parameters. In case of
the PID controller, control parameters (proportional gain $K_{p},$
integral gain $K_{i},$ and differential gain $K_{D}$) are obtained
offline before starting the closed-loop control operation. The TS
fuzzy controller consists of five rules, where univariate Gaussian
membership functions are utilized in each rule. To obtain the antecedents
and consequent parameters of the rules, the fuzzy controller is trained
with the PID controller's input and output datasets. It is clear that
in both PID and TS fuzzy controllers the parameters are fixed before
the starting of the closed-loop control system. On the contrary, in
G-controller not only the GENEFIS but also the parameters of the sliding
surface are evolving. Those sliding parameters are initialized with
a very small value, then evolved to a desired value by using different
learning rates. These rates are varied with respect to the corresponding
plants, and desired actions. To the best of our knowledge, this approach
of evolving the sliding parameters is never utilized before in any
of the existing evolving neuro-fuzzy controllers. It makes the proposed
G-controller a fully self-evolving controller.

\begin{center}
\begin{figure}[h]
\begin{centering}
\includegraphics[scale=0.17]{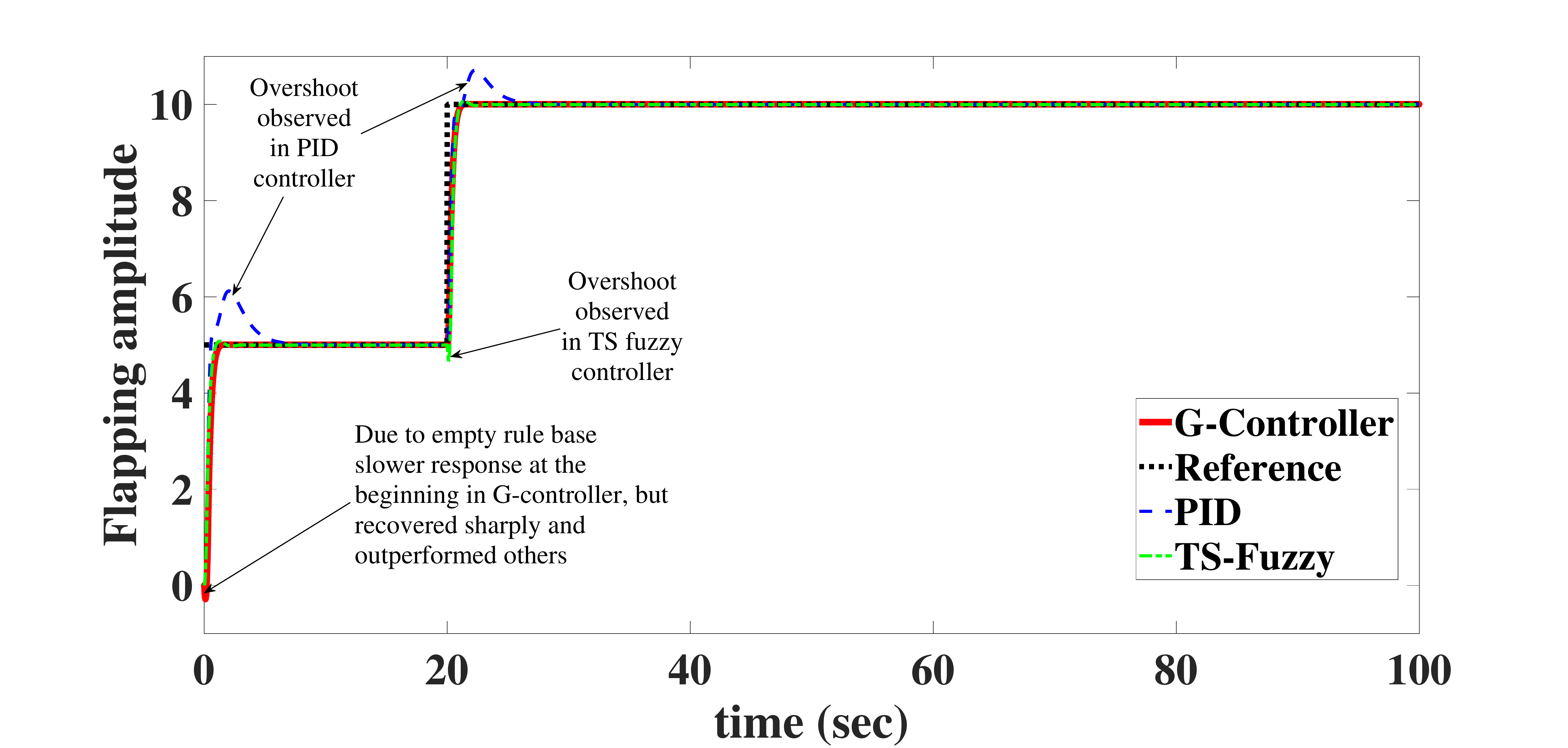}
\par\end{centering}

\caption{Performance observation of various controllers in altitude tracking
of BIFW MAV when the trajectory is a step function $Z_{d}(t)=5u(t)+5u(t-20)$
\label{fig:step-fw}}
\end{figure}

\par\end{center}

\begin{center}
\begin{figure}[th]
\begin{centering}
\includegraphics[scale=0.17]{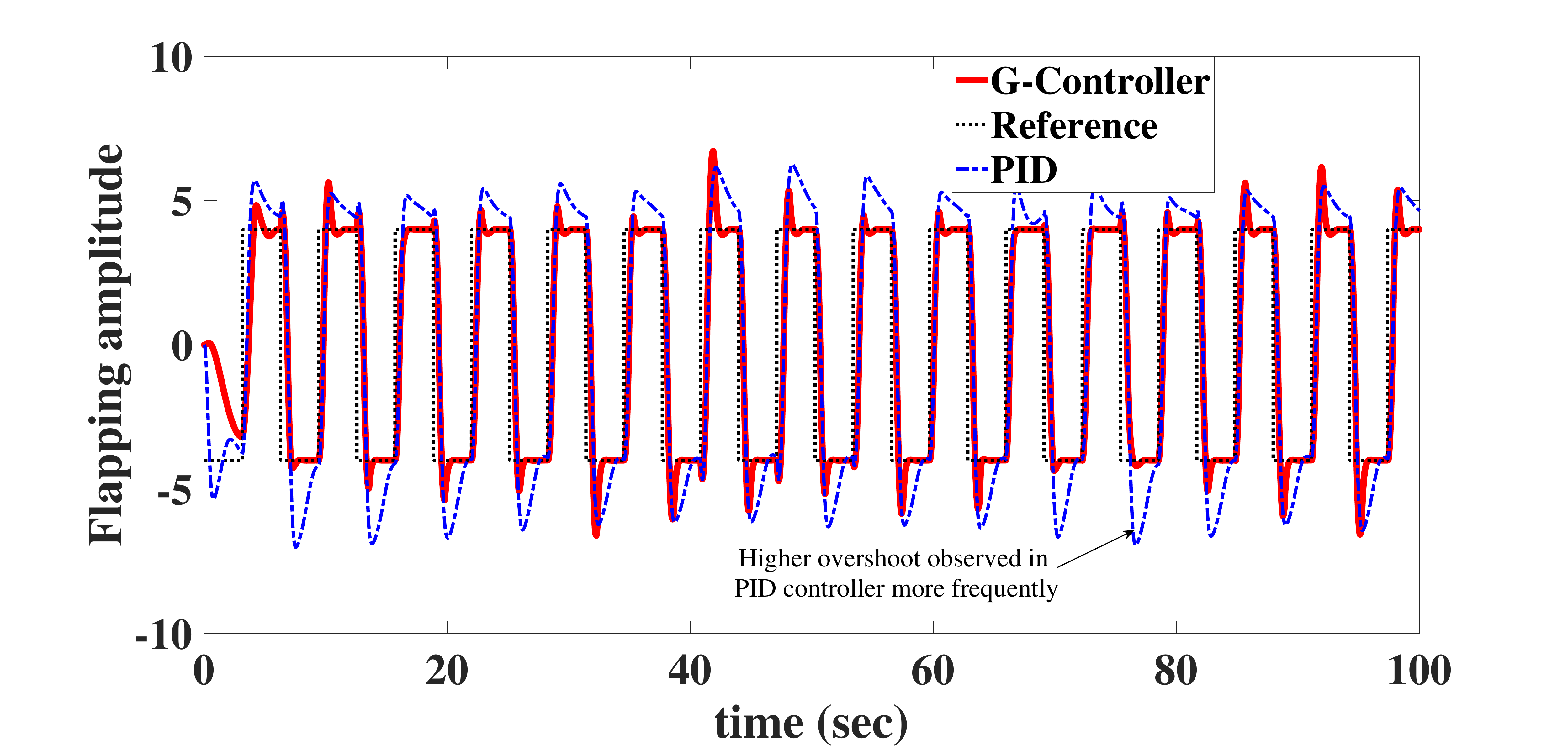}
\par\end{centering}

\caption{Performance observation of various controllers in altitude tracking
of BIFW MAV when the trajectory is a square wave function with an
amplitude of 4 m and frequency 1 Hz\label{fig:square_f1_a4_FW}}
\end{figure}

\par\end{center}

\begin{center}
\begin{figure}[th]
\begin{centering}
\includegraphics[scale=0.17]{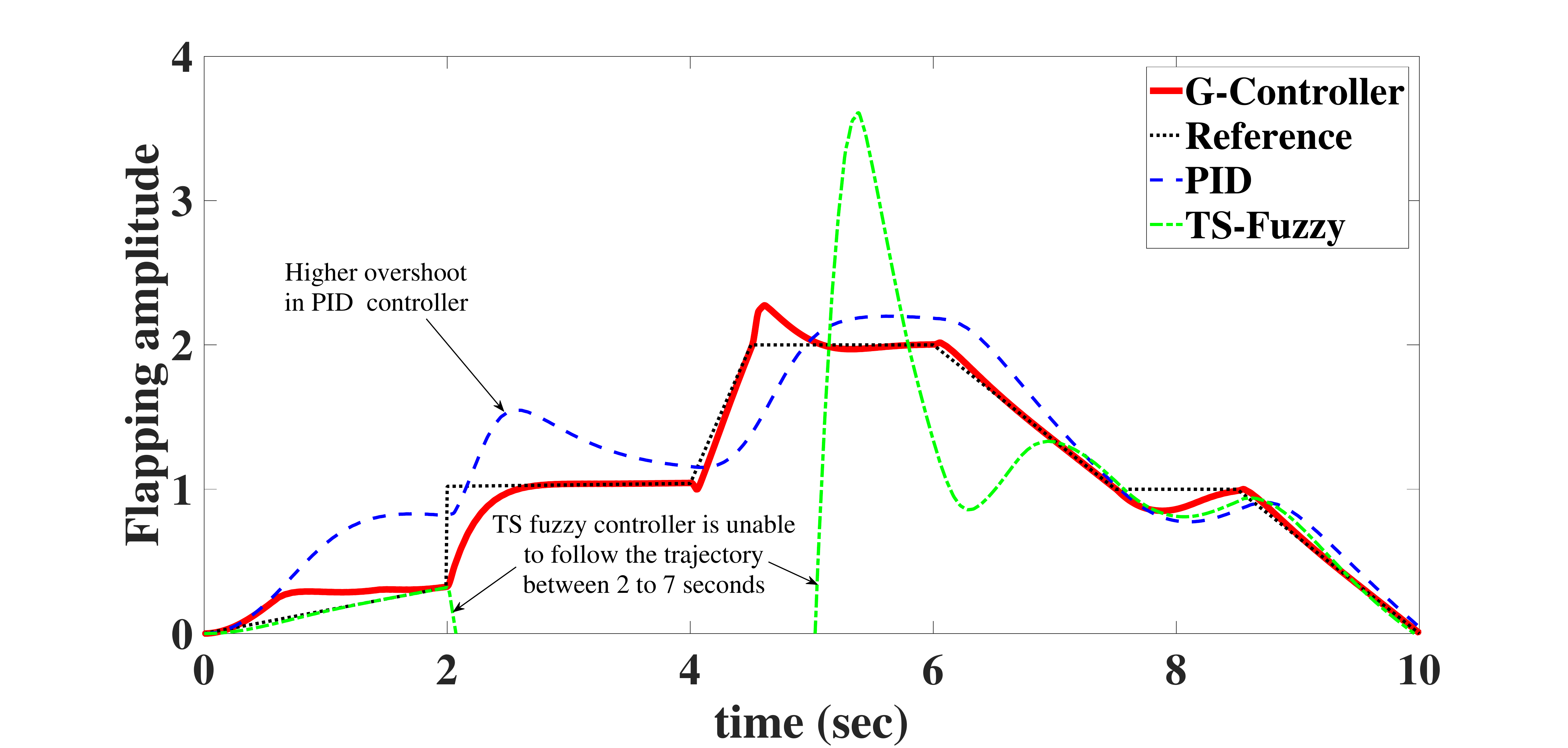}
\par\end{centering}

\caption{Performance observation of various controllers in BIFW MAV in case
of tracking a customized trajectory\label{fig:Sig-build}}
\end{figure}

\par\end{center}

\begin{center}
\begin{figure}[th]
\begin{centering}
\includegraphics[scale=0.17]{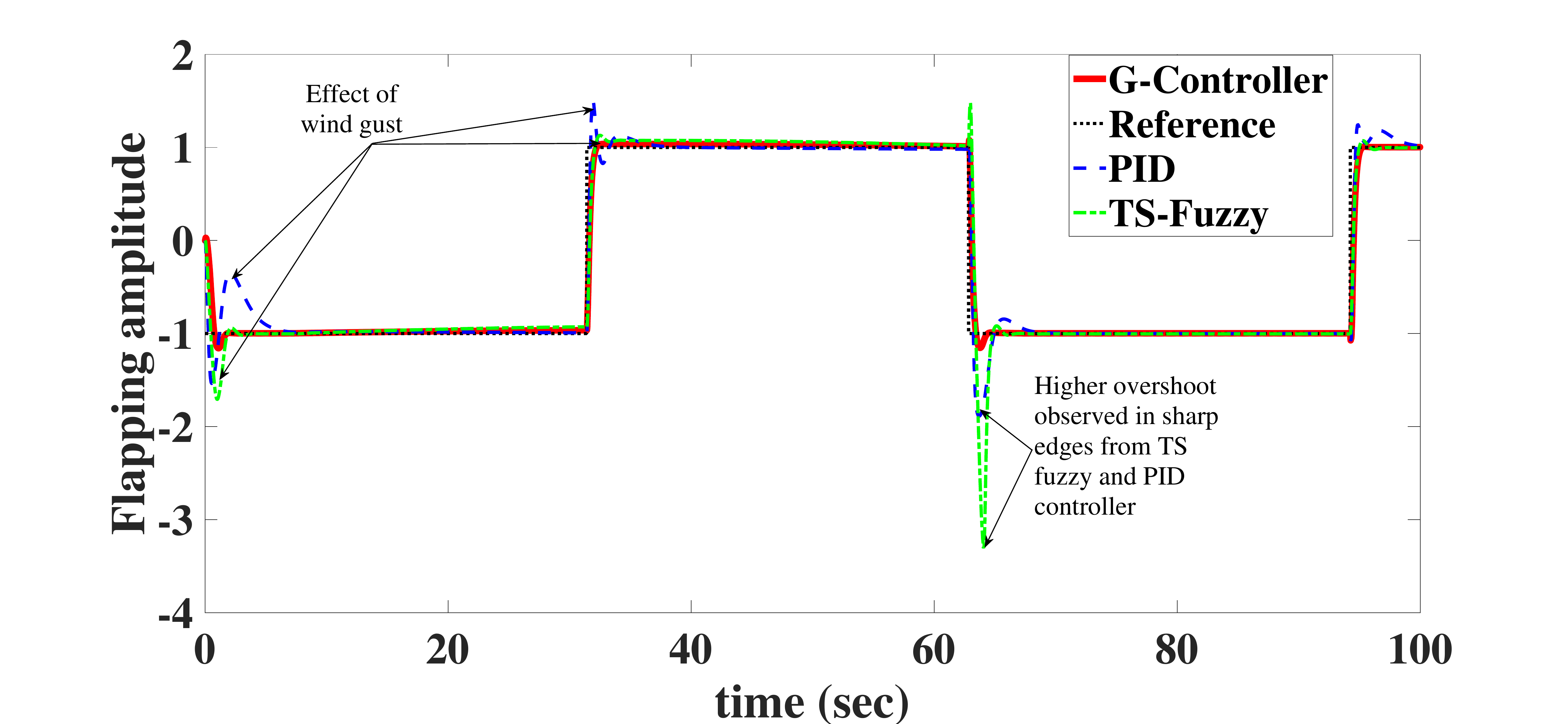}
\par\end{centering}

\caption{Performance observation of various controllers in altitude tracking
of BIFW MAV by considering wind gust as environmental uncertainty\label{fig:gust}}

\end{figure}

\par\end{center}

\begin{center}
\begin{table*}[t]
\caption{Measured RMSE, rise and settling time of various controllers in operating
the BIFW MAV}
\label{tab:rmse}

\centering{}%
\begin{tabular}{|>{\raggedright}p{2.35cm}|>{\raggedright}m{1.45cm}|>{\raggedright}m{2.2cm}|>{\raggedright}m{1cm}|>{\raggedright}m{1.1cm}|>{\raggedright}m{1.4cm}|>{\raggedright}m{1.3cm}|>{\raggedright}m{1.3cm}|>{\raggedright}m{1.4cm}|}
\hline 
\multirow{2}{2.35cm}{\textbf{Desired Trajectory ($Z_{d}$)}} & \multirow{2}{1.45cm}{\textbf{Maximum amplitude (meter)}} & \multirow{2}{2.2cm}{\textbf{Measured features}} & \multicolumn{3}{c|}{\textbf{Without gust}} & \multicolumn{3}{c|}{\textbf{With gust}}\tabularnewline
\cline{4-9} 
 &  &  & \textbf{PID} & \textbf{TS-}

\textbf{Fuzzy} & \textbf{G-controller} & \textbf{PID} & \textbf{TS-}

\textbf{Fuzzy} & \textbf{G-controller}\tabularnewline
\hline 
\multirow{3}{2.35cm}{\textbf{Constant height}} & \multirow{3}{1.45cm}{10} & \textbf{RMSE} & 0.6630 & \textbf{0.5693} & 0.6460 & 0.6635 & \textbf{0.5708} & 0.6461\tabularnewline
\cline{3-9} 
 &  & \textbf{Rise time (sec)} & 0.9040 & \textbf{0.7442} & 1.2583 & 0.9040 & \textbf{0.7442} & 0.7984\tabularnewline
\cline{3-9} 
 &  & \textbf{Settling time (sec)} & 8.2595 & 4.9305 & \textbf{3.9525} & 10.1 & 3.6 & \textbf{2.6}\tabularnewline
\hline 
\multirow{9}{2.35cm}{\textbf{Step function}} & \multirow{3}{1.45cm}{$Z_{d}(t)=5u(t)+5u(t-20)$} & \textbf{RMSE} & 0.4000 & 0.4023 & \textbf{0.3866} & 0.3962 & 0.4002 & \textbf{0.3865}\tabularnewline
\cline{3-9} 
 &  & \textbf{Rise time (sec)} & 21.242 & \textbf{21.094} & 21.37 & 21.23 & \textbf{21.05} & 21.12\tabularnewline
\cline{3-9} 
 &  & \textbf{Settling time (sec)} & 28.1 & 23.01 & \textbf{21.50} & 27.5 & 23.6 & \textbf{21.80}\tabularnewline
\cline{2-9} 
 & \multirow{3}{1.45cm}{$Z_{d}(t)=10u(t-20)$} & \textbf{RMSE} & 0.6266 & 0.6701 & \textbf{0.5677} & 0.6216 & 0.6616 & \textbf{0.5611}\tabularnewline
\cline{3-9} 
 &  & \textbf{Rise time (sec)} & 21.03 & 20.91 & \textbf{20.746} & 21.02 & 20.81 & \textbf{20.75}\tabularnewline
\cline{3-9} 
 &  & \textbf{Settling time (sec)} & 29.50 & \textbf{22.05} & 22.07 & 32.02 & 24.99 & \textbf{21.66}\tabularnewline
\cline{2-9} 
 & \multirow{3}{1.45cm}{$Z_{d}(t)=-5u(t)+10u(t-20)$} & \textbf{RMSE} & 0.6695 & 0.7188 & \textbf{0.6535} & 0.6648 & 0.7109 & \textbf{0.6477}\tabularnewline
\cline{3-9} 
 &  & \textbf{Rise time (sec)} & 21.035 & \textbf{20.95} & 21.53 & 21.02 & 20.81 & \textbf{20.79}\tabularnewline
\cline{3-9} 
 &  & \textbf{Settling time (sec)} & 29.5 & 22.5 & \textbf{22.1} & 31.56 & 24.59 & \textbf{21.03}\tabularnewline
\hline 
\multirow{9}{2.35cm}{\textbf{Square wave function ($f=0.1\; Hz$)}} & \multirow{3}{1.45cm}{1} & \textbf{RMSE} & 0.2039 & 0.2493 & \textbf{0.1746} & 0.2020 & 0.2552 & \textbf{0.1710}\tabularnewline
\cline{3-9} 
 &  & \textbf{Rise time (sec)} & \textbf{0.38} & 0.48 & 0.83 & \textbf{63.18} & 63.43 & 63.45\tabularnewline
\cline{3-9} 
 &  & \textbf{Settling time (sec)} & 10.11 & 3.8 & \textbf{2.75} & 69.52 & 69.50 & \textbf{67.05}\tabularnewline
\cline{2-9} 
 & \multirow{3}{1.45cm}{4} & \textbf{RMSE} & 0.8909 & 0.9059 & \textbf{0.8337} & 0.8892 & 0.9025 & \textbf{0.8311}\tabularnewline
\cline{3-9} 
 &  & \textbf{Rise time (sec)} & \textbf{0.45} & 0.524 & 0.914 & \textbf{0.49} & 0.52 & 0.91\tabularnewline
\cline{3-9} 
 &  & \textbf{Settling time (sec)} & 6.75 & 5.6 & \textbf{2.5} & 8.92 & 5.14 & \textbf{2.15}\tabularnewline
\cline{2-9} 
 & \multirow{3}{1.45cm}{10} & \textbf{RMSE} & 2.7381 & 511.9071 & \textbf{2.5406} & 2.7339 & 1629.4883 & \textbf{2.5376}\tabularnewline
\cline{3-9} 
 &  & \textbf{Rise time (sec)} & \textbf{23.30} & N/A & 23.37 & \textbf{0.97} & N/A & 1.26\tabularnewline
\cline{3-9} 
 &  & \textbf{Settling time (sec)} & 42.5 & N/A & \textbf{34.5} & 8.37 & N/A & \textbf{2.25}\tabularnewline
\hline 
\multirow{6}{2.35cm}{\textbf{Square wave function($f=1\; Hz$)}} & \multirow{3}{1.45cm}{1} & \textbf{RMSE} & 0.6719 & 362.9960 & \textbf{0.5893} & 0.6720 & 363.3044 & \textbf{0.5735}\tabularnewline
\cline{3-9} 
 &  & \textbf{Rise time (sec)} & \textbf{3.43} & N/A & 3.72 & \textbf{3.435} & N/A & 4.17\tabularnewline
\cline{3-9} 
 &  & \textbf{Settling time (sec)} & 22.6 & N/A & \textbf{8.5} & 18.1 & N/A & \textbf{4.4}\tabularnewline
\cline{2-9} 
 & \multirow{3}{1.45cm}{4} & \textbf{RMSE} & 3.1435 & 573.3459 & \textbf{2.7275} & 3.1336 & 1258.0926 & \textbf{2.6766}\tabularnewline
\cline{3-9} 
 &  & \textbf{Rise time (sec)} & \textbf{3.752} & N/A & 4.06 & \textbf{4.13} & N/A & 4.62\tabularnewline
\cline{3-9} 
 &  & \textbf{Settling time (sec)} & 12.4 & N/A & \textbf{7.81} & 6.58 & N/A & \textbf{4.74}\tabularnewline
\hline 
\multirow{3}{2.35cm}{\textbf{Customized wave function}} & \multirow{3}{1.45cm}{2} & \textbf{RMSE} & 0.2856 & 4.7793 & \textbf{0.1033} & 0.2846 & 4.7708 & \textbf{0.1026}\tabularnewline
\cline{3-9} 
 &  & \textbf{Rise time (sec)} & 2.0 & \textbf{1.30} & 1.98 & 2.16 & \textbf{0.93} & 1.96\tabularnewline
\cline{3-9} 
 &  & \textbf{Settling time (sec)} & 5.5 & 7.25 & \textbf{2.8} & 7.66 & 7.25 & \textbf{2.75}\tabularnewline
\hline 
\multirow{3}{2.35cm}{\textbf{Sawtooth wave function}} & \multirow{3}{1.45cm}{1} & \textbf{RMSE} & 0.5235 & 325.4397 & \textbf{0.4781} & 0.5240 & 325.4111 & \textbf{0.4776}\tabularnewline
\cline{3-9} 
 &  & \textbf{Rise time (sec)} & \textbf{0.24} & N/A & 0.96 & 0.9039 & N/A & \textbf{0.6499}\tabularnewline
\cline{3-9} 
 &  & \textbf{Settling time (sec)} & 4.1 & N/A & \textbf{1.5} & 5.68 & N/A & \textbf{0.65}\tabularnewline
\hline 
\multirow{3}{2.35cm}{\textbf{Sine wave function}} & \multirow{3}{1.45cm}{1} & \textbf{RMSE} & 0.2096 & 0.0737 & \textbf{0.0356} & 0.1880 & 0.0764 & \textbf{0.0395}\tabularnewline
\cline{3-9} 
 &  & \textbf{Rise time (sec)} & \textbf{0.824} & 1.19 & 1.01 & \textbf{0.7405} & 1.19 & 1.08\tabularnewline
\cline{3-9} 
 &  & \textbf{Settling time (sec)} & 4.1 & 3.4 & \textbf{2.7} & 5.23 & 3.32 & \textbf{2.11}\tabularnewline
\hline 
\end{tabular}
\end{table*}

\par\end{center}

\begin{center}
\begin{figure}[th]
\begin{centering}
\includegraphics[scale=0.15]{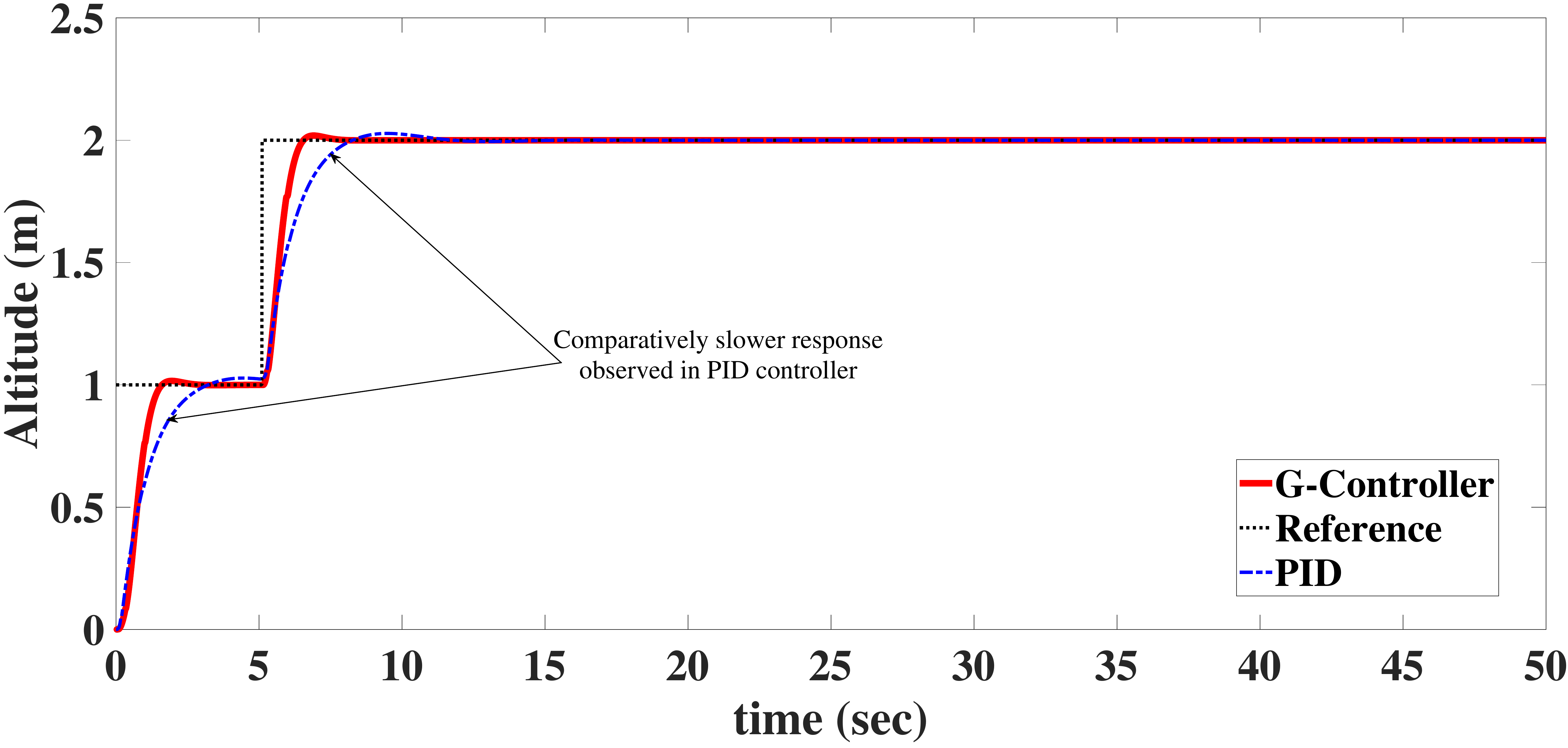}
\par\end{centering}

\caption{Performance observation of a PID and proposed G-controller in tracking
various altitude of a hexacopter\label{fig:altitude_hexa}}
\end{figure}

\par\end{center}

\begin{figure}[th]
\begin{centering}
\subfloat[]{\begin{centering}
\includegraphics[scale=0.15]{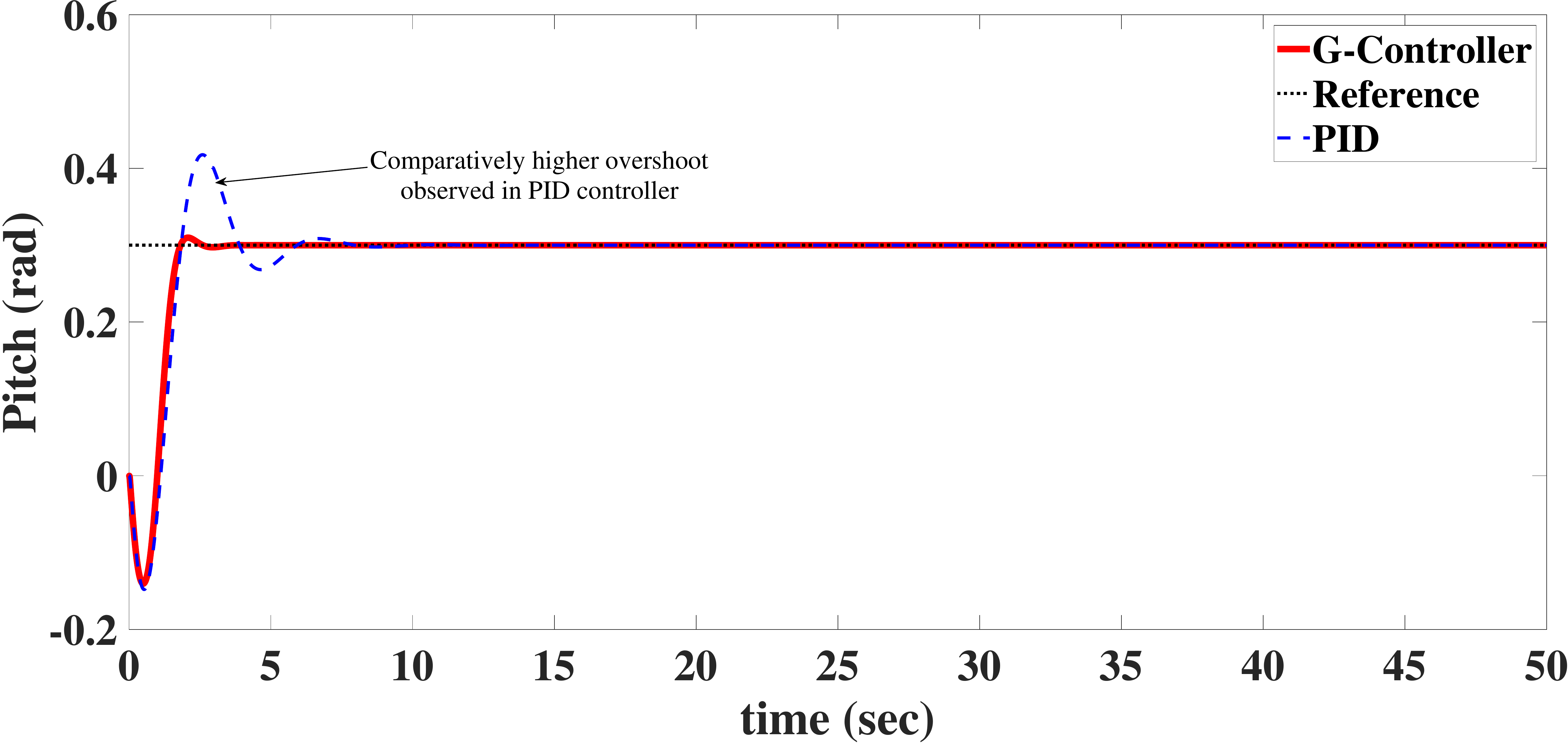}
\par\end{centering}

}
\par\end{centering}

\begin{centering}
\subfloat[]{\begin{centering}
\includegraphics[scale=0.15]{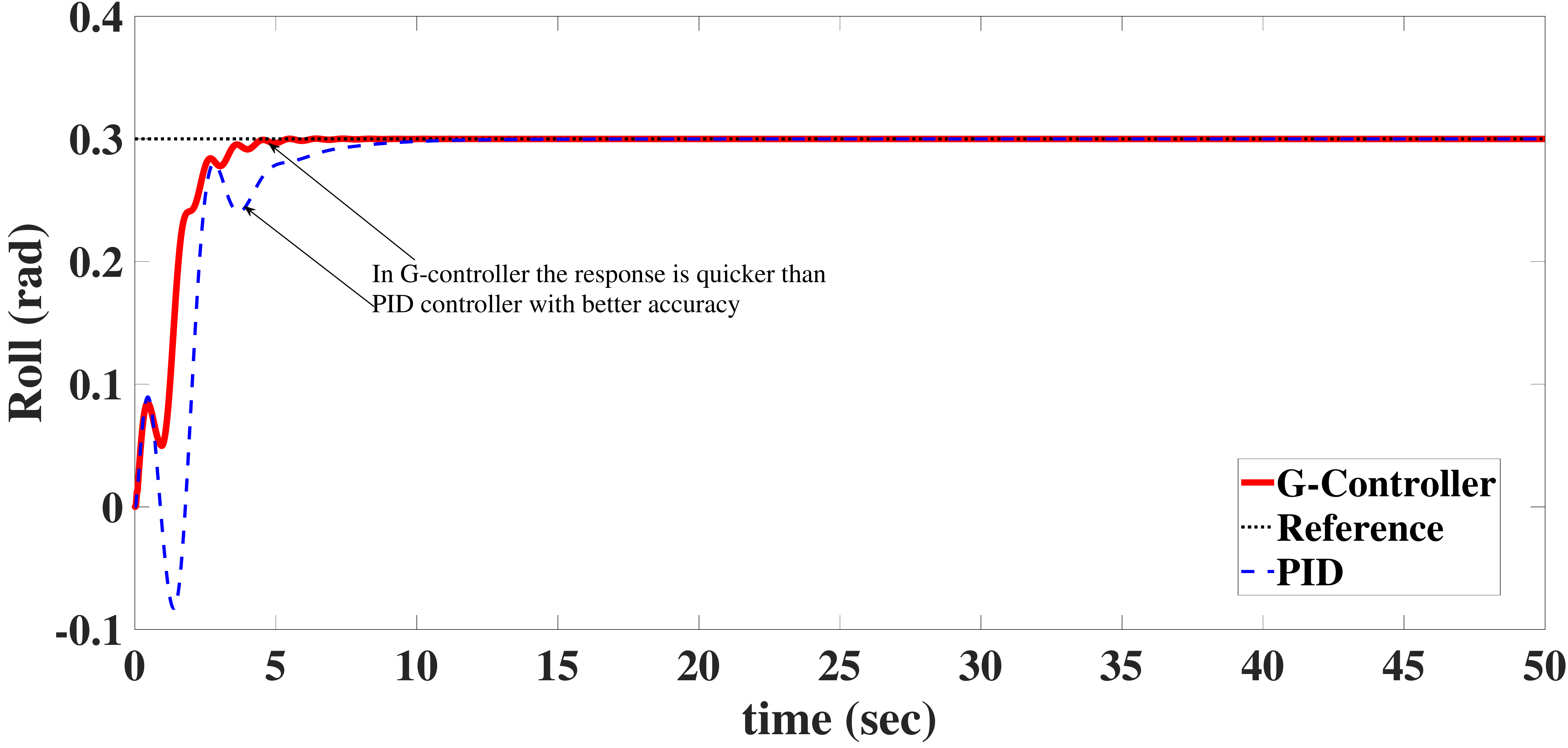}
\par\end{centering}

}
\par\end{centering}

\caption{Performance observation of a PID and proposed G-controller in tracking
desired (a) pitching and (b) rolling motion of a hexacopter\label{fig:roll_pitch_hexa}}
\end{figure}

\begin{figure}[t]
\begin{centering}
\subfloat[]{\includegraphics[scale=0.08]{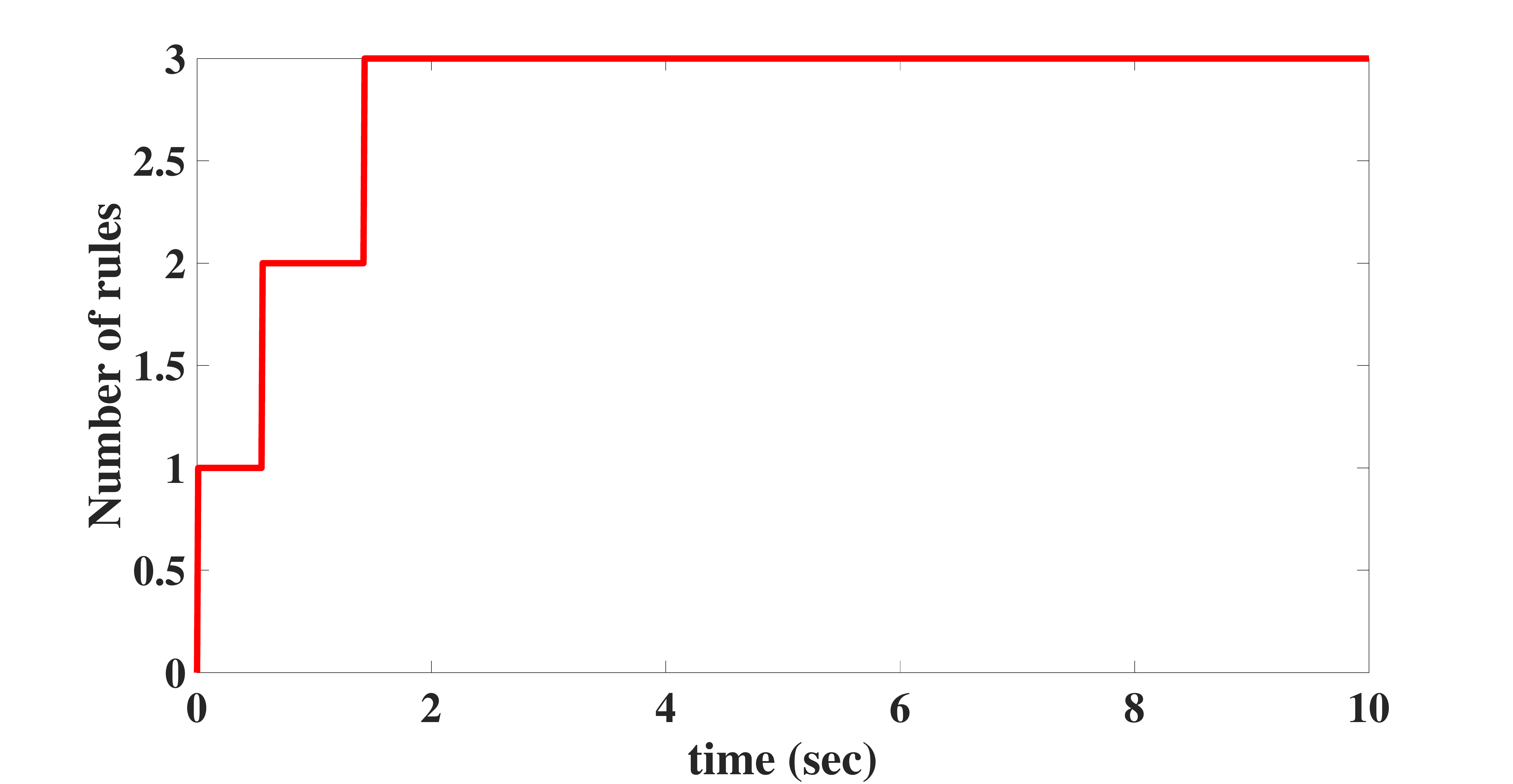}

}\subfloat[]{\includegraphics[scale=0.0837]{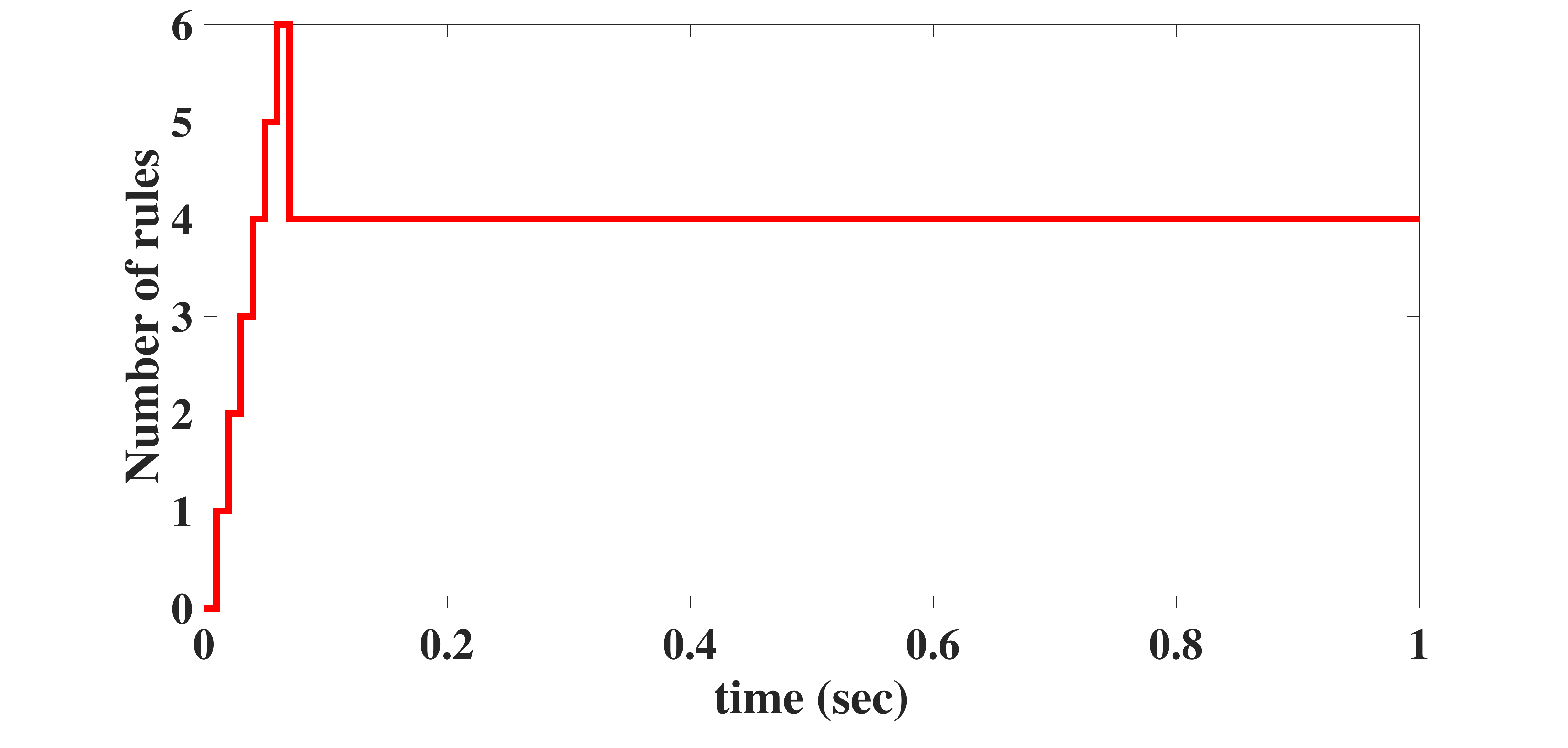}

}
\par\end{centering}

\begin{centering}
\subfloat[]{\includegraphics[scale=0.08]{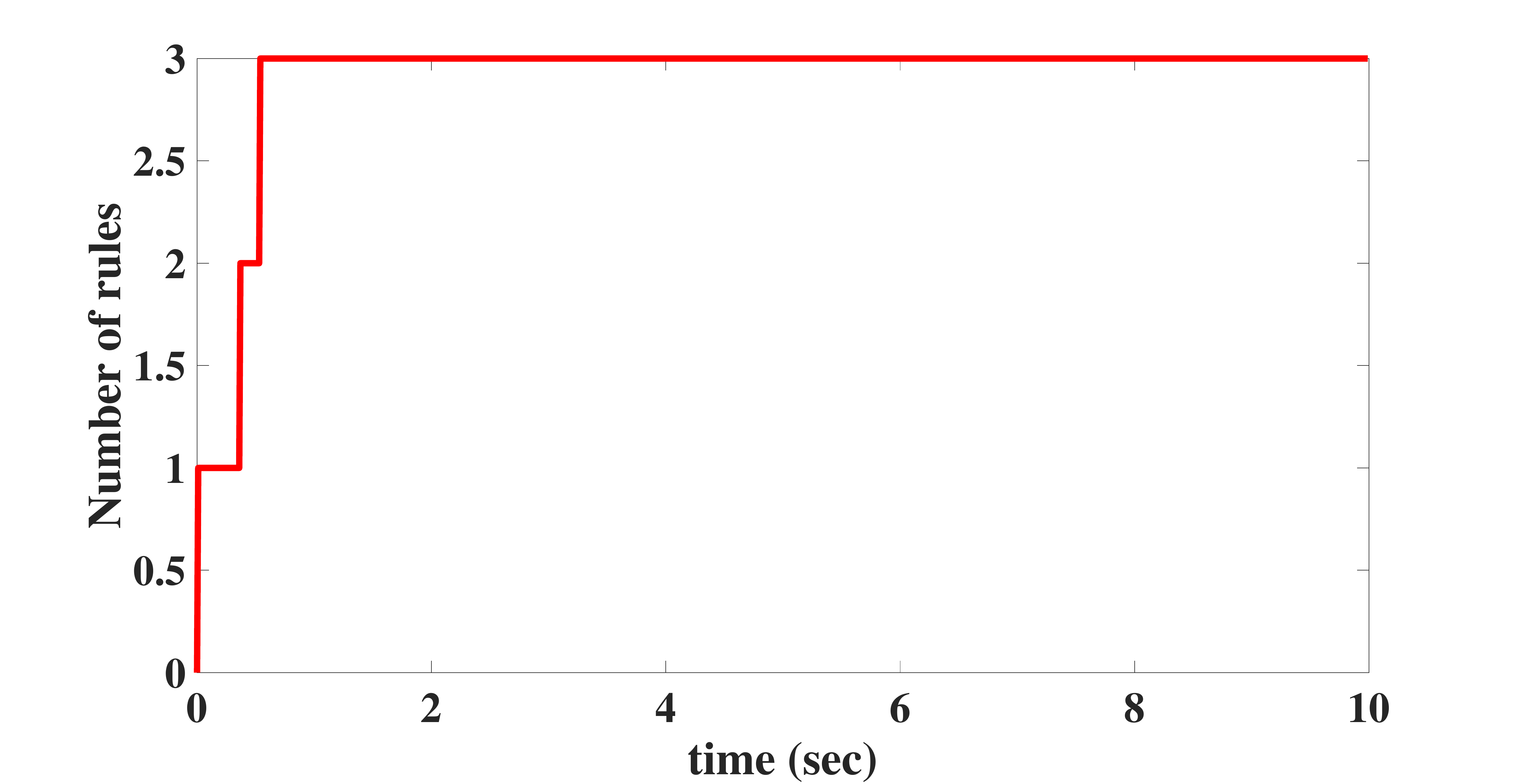}

}\subfloat[]{\includegraphics[scale=0.0827]{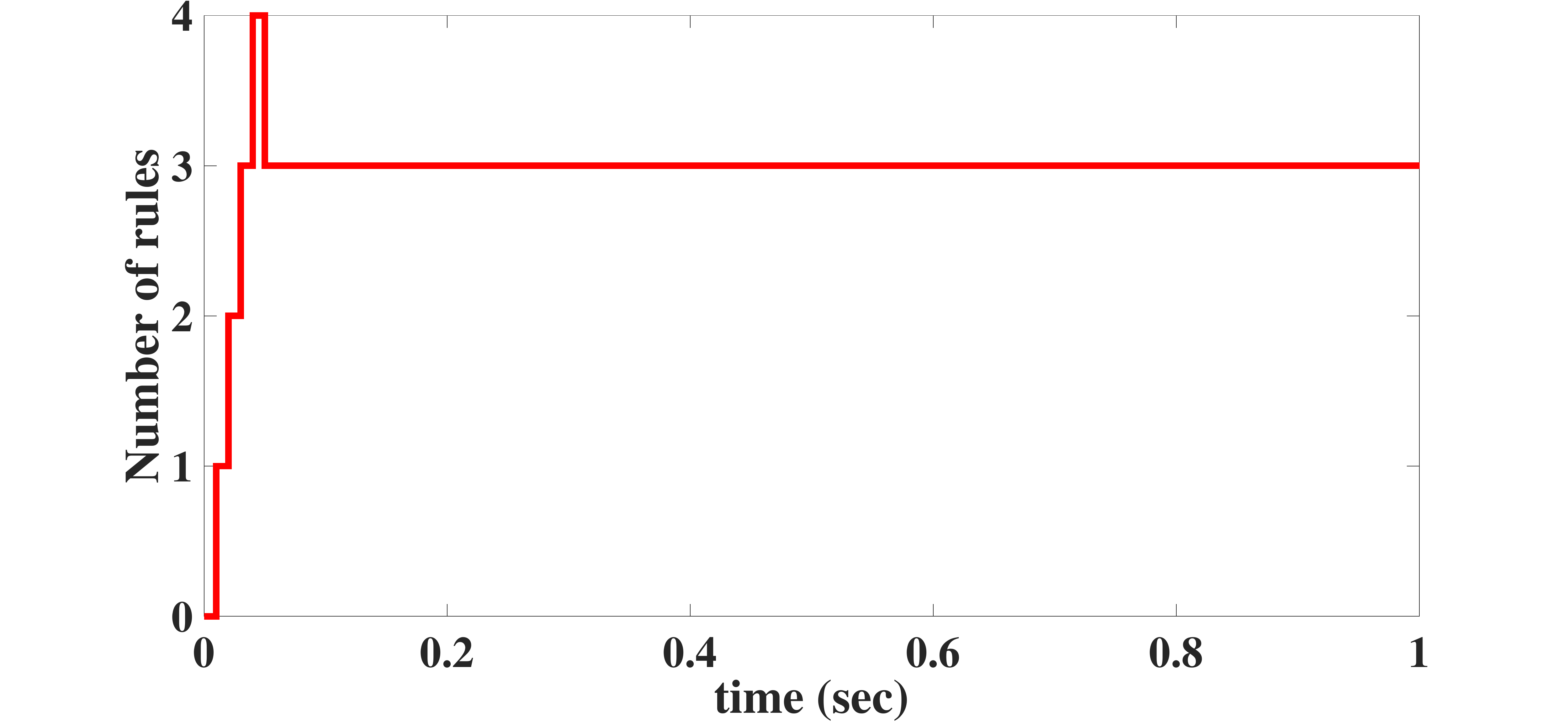}

}
\par\end{centering}

\caption{Generated rules of the self-evolving G-controller at various trajectories
of BIFW MAV and hexacopter where the trajectories are (a) step function
altitude for BIFW MAV, (b) customized altitude for BIFW MAV, (c) pitching
for Hexacopter, (d) rolling for Hexacopter\label{fig:Generated-rules}}
\end{figure}

\section{Conclusions\label{sec:Conclusions}}

The G-controller developed in this work is an entirely model-free
approach and self-evolving in nature, i.e. it can alter its structure,
and system parameters online to cope with changing dynamics of the
plant to be controlled. Besides, the synthesis of SMC theory based
adaptation laws improve its robustness against various internal and
external uncertainties. These desirable features make the G-controller
a suitable candidate for highly nonlinear autonomous vehicles. In
this work, our proposed control algorithm is developed using C programming
language considering the compatibility issues to implement directly
in hardware of a variety of autonomous vehicles like BIFW MAV, quadcopter,
hexacopter, etc. The controller's performance has been evaluated by
observing the tracking performance of an over-actuated BIFW MAV and
an over-actuated hexacopter's plant with respect to a variety of trajectories.
The performances are compared to that of a PID, and a TS fuzzy controller
to observe the improvements in our proposed online evolving G-controller.
The G-controller starts building the structure from scratch with an
empty fuzzy set in the closed-loop system. It causes a slow response
at the very beginning of the loop, which is a common phenomenon in
any self-evolving controller. However, due to the integration of GRAT+,
multivariate Gaussian function, SMC learning theory based adaptation
laws, the self-evolving mechanism of the G-controller is faster with
a lower computational cost. In addition, wind gust has been added
to the BIFW MAV plant as environmental uncertainties to evaluate the
G-controller's robustness against unknown perturbations, where satisfactory
tracking of the desired trajectory proves the proposed controller
performance to eliminate various uncertainties. Thus, the G-controller's
stability is confirmed by both the Lyapunov theory and experiments.
In future, the controller will be executed through hardware-based
flight test of various unmanned aerial vehicles.

\appendices{}

\section*{Acknowledgement}

The authors would like to thank the Australian Defense Science and
Technology Group for providing the simulated BIFW MAV plant, Unmanned
Aerial Vehicle laboratory of the UNSW at the Australian Defense Force
Academy for supporting with the hexacopter plants, and the computational
support from the Computational Intelligence Laboratory of Nanyang
Technological University (NTU) Singapore. This work is fully supported
by NTU start-up grant and MOE tier-1 grant.

\bibliographystyle{IEEEtran}
\bibliography{g_controller}

\begin{IEEEbiography}[{{\includegraphics[clip,width=1in,height=1.25in,keepaspectratio,bb = 0 0 200 100, draft, type=eps]{C:/Users/N1705348C/examples/CV-image}}}]
{Md Meftahul Ferdaus} All about you and the what your interests
are.
\end{IEEEbiography}

\begin{IEEEbiographynophoto}
{Mahardhika Pratama} M\end{IEEEbiographynophoto}

\end{document}